\documentclass[12pt]{article}
\usepackage[dvips]{graphicx}
\usepackage{epsfig}
\usepackage{amsmath,amssymb,amsbsy}
\usepackage{amsfonts}
\begin{document}
\thispagestyle{empty}
\begin{center}

{\Large\bf{Semiinclusive DIS cross sections and spin asymmetries
in the quantum statistical parton distributions approach}}

\vskip1.4cm
{\bf Claude Bourrely}
\vskip 0.3cm
D\'epartement de Physique, Facult\'e des Sciences de Luminy,\\
Universit\'e de la M\'editerran\'ee - Aix-Marseille II,\\
13288 Marseille, Cedex 09, France\\
\vskip 0.5cm
{\bf Franco Buccella}
\vskip 0.3cm
Dipartimento di Scienze Fisiche, Universit\`a di Napoli,\\
Via Cintia, I-80126, Napoli
and INFN, Sezione di Napoli, Italy
\vskip 0.5cm
{\bf Jacques Soffer}
\vskip 0.3cm
Physics Department, Temple University\\
Barton Hall, 1900 N, 13th Street\\
Philadelphia, PA 19122-6082, USA
\vskip 0.5cm
{\bf Abstract}\end{center}

We consider the extension of the statistical parton distributions to include
their transverse momentum dependence and
we take into account the effects of the Melosh-Wigner rotation for the
polarized distributions. With a suitable choice of the fragmentation function, we make predictions for the recent semiinclusive Deep Inelastic Scattering data on the cross section
and double longitudinal-spin asymmetries from JLab.
We also give some predictions for future experiments on electron-neutron 
scattering.

\vskip 0.5cm


\vskip 0.5cm

\noindent UNIV. NAPLES DSF/8/2010
\newpage
\section{Introduction}
 A new set of parton distribution functions (PDF) was constructed in the
framework of a statistical approach of the nucleon \cite{bbs1}, and let us
first recall very briefly, its main characteristic features. For quarks
(antiquarks), the building blocks are the helicity dependent distributions
$q_{\pm}(x)$ ($\bar q_{\pm}(x)$) so this allows to describe simultaneously the
unpolarized distributions $q(x)= q_{+}(x)+q_{-}(x)$ and the helicity
distributions $\Delta q(x) = q_{+}(x)-q_{-}(x)$ (similarly for antiquarks). At
the initial energy scale taken at $Q^2_0= 4 \mbox{GeV}^2$, these distributions
are given by the sum of two terms, a quasi Fermi-Dirac function and a helicity
independent diffractive
contribution, which leads to a universal behavior at very low $x$ for all
flavors. The flavor asymmetry for the light sea, {\it i.e.} $\bar d (x) > \bar
u (x)$, observed in the data is built in. This is clearly understood in terms
of the Pauli exclusion principle, based on the fact that the proton contains
two $u$ quarks and only one $d$ quark. The chiral properties of QCD lead to
strong relations between $q(x)$ and $\bar q (x)$.
For example, it is found that the well estalished result $\Delta u (x)>0 $\
implies $\Delta
\bar u (x)>0$ and similarly $\Delta d (x)<0$ leads to $\Delta \bar d (x)<0$.
Concerning the gluon, the unpolarized distribution $G(x,Q_0^2)$ is given in
terms of a quasi Bose-Einstein function, with only {\it one free parameter},
and for simplicity, one assumes zero gluon polarization, {\it i.e.} $\Delta
G(x,Q_0^2)=0$, at the initial energy scale $Q_0^2$. All unpolarized and
polarized light quark distributions depend upon {\it eight}
free parameters, which were determined in 2002 (see Ref.~\cite{bbs1}), from a
next-to-leading order fit of a selected set of accurate DIS data. Concerning
the strange quarks and antiquarks distributions, the statistical approach has
been applied to calculate the strange quark asymmetry and the corresponding
helicity distributions, which were found both negative at all $x$ values
\cite{bbs2}. More recently, new tests against experimental (unpolarized and
polarized) data turned out to be very satisfactory, in particular in hadronic
reactions, as reported in Refs.~\cite{bbs3,bbs4}.\\
The paper is organized as follows. In the next section we review the
construction of the statistical distributions and we present an improved
version
of the extension to the transverse momentum dependence (TMD).
 In section 3, we will consider charged pion production in
semiinclusive deep inelastic scattering (SIDIS), $\ell \, N \to \ell \, H \,
X$, a suitable reaction for testing our TMD distributions, more specifically,
for the cross section and the longitudinal-spin asymmetry, by taking into
account the effects of the Melosh-Wigner rotation. The results are given and
discussed in section 4 and the last section is devoted to our concluding
remarks.

\section{The TMD parton distributions}
\subsection{The original longitudinal parton distributions}
We now review some of the basic features of the statistical approach, as oppose
to the standard polynomial type
parametrizations of the PDF, based on Regge theory at low $x$ and counting
rules at large $x$.
The fermion distributions are given by the sum of two terms \cite{bbs1},
a quasi Fermi-Dirac function and a helicity independent diffractive
contribution equal for all light quarks:
\begin{equation}
xq^h(x,Q^2_0)=
\frac{AX^h_{0q}x^b}{\exp [(x-X^h_{0q})/\bar{x}]+1}+
\frac{\tilde{A}x^{\tilde{b}}}{\exp(x/\bar{x})+1}~,
\label{eq1}
\end{equation}
\begin{equation}
x\bar{q}^h(x,Q^2_0)=
\frac{{\bar A}(X^{-h}_{0q})^{-1}x^{2b}}{\exp [(x+X^{-h}_{0q})/\bar{x}]+1}+
\frac{\tilde{A}x^{\tilde{b}}}{\exp(x/\bar{x})+1}~,
\label{eq2}
\end{equation}
at the input energy scale $Q_0^2=4 \mbox{GeV}^2$.\\
Notice the change of sign of the potentials
and helicity for the antiquarks.\\
The parameter $\bar{x}$ plays the role of a {\it universal temperature}
and $X^{\pm}_{0q}$ are the two {\it thermodynamical potentials} of the quark
$q$, with helicity $h=\pm$.
The {\it eight} free parameters\footnote{$A=1.74938$ and $\bar{A}~=1.90801$ are
fixed by the following normalization conditions $u-\bar{u}=2$, $d-\bar{d}=1$.}
in Eqs.~(\ref{eq1},\ref{eq2}) were
determined at the input scale from the comparison with a selected set of
very precise unpolarized and polarized DIS data \cite{bbs1}. They have the
following values
\begin{equation}
\bar{x}=0.09907,~ b=0.40962,~\tilde{b}=-0.25347,~\tilde{A}=0.08318,
\label{eq3}
\end{equation}
\begin{equation}
X^+_{0u}=0.46128,~X^-_{0u}=0.29766,~X^-_{0d}=0.30174,~X^+_{0d}=0.22775~.
\label{eq4}
\end{equation}
For the gluons we consider the black-body inspired expression
\begin{equation}
xG(x,Q^2_0)=
\frac{A_Gx^{b_G}}{\exp(x/\bar{x})-1}~,
\label{eq5}
\end{equation}
a quasi Bose-Einstein function, with $b_G=0.90$, the only free parameter
\footnote{In Ref.~\cite{bbs1} we were assuming that, for very small $x$,
$xG(x,Q^2_0)$ has the same behavior as $x\bar q(x,Q^2_0)$, so we took $b_G = 1
+ \tilde b$. However this choice leads to a too much rapid rise of the gluon
distribution, compared to its recent  determination from HERA data, which
requires $b_G=0.90$.}, since $A_G=20.53$ is determined by the momentum sum
rule.
 We also assume that, at the input energy scale, the polarized gluon,
distribution vanishes, so
\begin{equation}
x\Delta G(x,Q^2_0)=0~.
\label{eq6}
\end{equation}
For the strange quark distributions, the simple choice made in Ref. \cite{bbs1}
was greatly improved in Ref. \cite{bbs2}, but they will not be considered in
this paper.\\
In Eqs.~(\ref{eq1},\ref{eq2}) the multiplicative factors $X^{h}_{0q}$ and
$(X^{-h}_{0q})^{-1}$ in
the numerators of the non-diffractive parts of $q$'s and $\bar{q}$'s
distributions, imply a modification
of the quantum statistical form, we were led to propose in order to agree with
experimental data. The presence of these multiplicative factors was justified
in our earlier attempt to generate the TMD
\cite{bbs5}, as we will explain now, with a considerable improvement.

\subsection{The TMD statistical distributions revisited}
\label{tmdrev}
 Let us recall that the TMD of the non-diffractive part of a quark distribution
$q$ of helicity $h$ (first term in Eq.~(\ref{eq1})) was introduced by the
following multiplicative term
\begin{equation}
\frac{1}{\mbox{exp}[(k_T^2/x\mu^2 - Y^h_{0q})/\bar x] + 1}~,
\label{eq7}
\end{equation}
where $Y^h_{0q}$ is the thermodynamical potential associated to the quark
transverse momentum $k_T$ and $1/\mu^2$ is a Lagrange multiplier, whose value
is determined by a transverse energy sum rule.\\
We notice that this term induces a non 
factorizable $x$ and
$k_T$ dependence as it is assumed in some other parametrizations. In order to recover the original $x$ 
distributions, the integration of (\ref{eq7}) over $k_T^2$ gives
the explicit factor
\begin{equation}
\int_0^\infty \frac{dk^2_T}{\exp[(k^2_T/x\mu^2-Y^{h}_{0q})/\bar{x}]+1} = -
x \mu^2\bar{x} \mbox{Li}_1(-\exp[Y^{h}_{0q}/\bar{x}])~.
\label{eq8}
\end{equation}
Here $\mbox{Li}_1$ denotes the polylogarithm function of order 1, which is
known to arise from the integral of Fermi-Dirac distributions and
is such that
\begin{equation}
-\mbox{Li}_1(-\mbox{e}^y)= \int_0^\infty
\frac{d\omega}{\mbox{e}^{(\omega-y)}+1}=
\ln{(1+\mbox{e}^{y})}~.
\label{eq9}
\end{equation}
Similarly for an antiquark  distribution $\bar q$ of helicity $-h$, according
to the rules of the statistical approach, one should use the same potential
with opposite sign, so one gets instead, $x \mu^2\bar{x}
\mbox{ln}(1+\exp[-Y^h_{0q}/\bar x])$. In Ref.~\cite{bbs5}, we made the
arbitrary simple choice $Y^h_{0q}=kX^h_{0q}$, with $k=1.42$, which allows to
recover the factor $X^h_{0q}$ for quarks (see Eq.~(\ref{eq1})), since for large
values \footnote{$\bar x$ has a small value according to Eq.~(\ref{eq3})
above.} of $Y^{h}_{0q}/\bar{x}$, one has
$\mbox{Li}_1(-\exp[Y^{h}_{0q}/\bar{x}])\sim Y^{h}_{0q}/\bar{x}$, which is
proportional to $X^{h}_{0q}$. However this is not suitable to get the factor
$[X^{h}_{0q}]^{-1}$ for antiquarks (see Eq.~(\ref{eq2})), because
$\mbox{Li}_1(-\exp[-Y^{h}_{0q}/\bar{x}])\sim \mbox{exp}[-Y^{h}_{0q}/\bar{x}]$
for large values of $Y^{h}_{0q}/\bar{x}$. In other words, the
 product $\mbox{ln}(1+\exp[Y^h_{0q}/\bar
x])\cdot\mbox{ln}(1+\exp[-Y^h_{0q}/\bar x])$ does not remain independent of
$Y^h_{0q}$, as it should.\\
 Actually, the division by $\bar x$ of the argument of the exponential in the
Fermi-Dirac expression was not necessary because for the transverse degrees of
freedom, $\mu^2$ plays the role of the temperature. This feature reflects the
fact that one should not treat on equal footing longitudinal and transverse
degrees of freedom. Therefore, for the sake of simplicity, we propose to
replace Eq.~(\ref{eq7}) by
\begin{equation}
\frac{1}{\mbox{exp}(k_T^2/x\mu^2 - Y^h_{0q}) + 1}~,
\label{eq10}
\end{equation}
with the corresponding integral over $k_T^2$, $x \mu^2
\mbox{ln}(1+\exp[Y^h_{0q}])$. Clearly this implies a different normalization
for $\mu^2$ and $Y^h_{0q}$. At high $k_T$, Eq.~(\ref{eq10}) has a Gaussian
behavior, with a width proportional to $\mu\sqrt{x}$, at variance with the
usual factorization assumption of the dependences in $x$ and $k_T$ \cite{aekp}.
The product $\mbox{ln}(1+\exp[Y^h_{0q}])\cdot\mbox{ln}(1+\exp[-Y^h_{0q}])$ has
its maximum $(\mbox{ln}2)^2$ for $Y^h_{0q}=0$ and therefore it is stationary
around this value.\\
So now in order to try to recover the factors $X^h_{0q}$ and
$(X^{-h}_{0q})^{-1}$ in Eqs.~(\ref{eq1},\ref{eq2}), we simply have to choose
$Y^h_{0q}$ such that $\mbox{ln}(1+\exp[Y^h_{0q}])$ is
proportional to $X^h_{0q}$ and more precisely such that
\begin{equation}
\mbox{ln}(1+\exp[Y^h_{0q}])=kX^h_{0q}~.
\label{eq11}
\end{equation}
This way we recover exactly the factors $X^h_{0q}$ introduced in
Eq.~(\ref{eq1}) for the quarks. We take the proportionality factor
$k = \ln{2}/X^+_{0d}$, $X^+_{0d}$ being the lowest longitudinal
potential, so with the value given in (\ref{eq4}) we get $k = 3.05$. In
order to get almost exactly $(X^h_{0q})^{-1}$ for the antiquarks in Eq.~(\ref{eq2}), we also assume
that the corresponding transverse potential $Y^+_{0d}$ is small and fixed to
the value 0.01. So from Eq.~(\ref{eq11}), the values of the other three transverse potentials can be obtained and we 
finally have
\begin{equation}
Y^+_{0u}=1.122,~Y^-_{0u}=0.388,~Y^-_{0d}=0.409,~Y^+_{0d}=0.010~.
\label{eq12}
\end{equation}
These are different from the values obtained in Ref.~\cite{bbs5} and will lead
to different predictions for the TMD of the PDF. The non-diffractive
contributions read now
\begin{equation}
xq^{h}(x,k_T^{2})=\frac{F(x)}{\exp(x-X^{h}_{0q})/\bar{x}+1}
\frac{1}{\exp(k^2_T/x\mu^2-Y^{h}_{0q})+1}~,
\label{nondifq}
\end{equation}
\begin{equation}
x\bar{q}^{h}(x,k_T^{2})=\frac{{\bar
F}(x)}{\exp(x+X^{-h}_{0q})/{\bar{x}}+1}
\frac{1}{\exp(k^2_T/x\mu^2+Y^{-h}_{0q})+1}~,
\label{nondifqbar}
\end{equation}
where
\begin{equation}
F(x) = \frac{A x^{b-1}
X^{h}_{0q}}{\mbox{ln}(1 + \exp{Y^{h}_{0q}})\mu^2}=\frac{A x^{b-1}}{k\mu^2}~.
\end{equation}
Similarly for $\bar q$ we have $\bar F(x)= \bar A x^{2b-1}/k\mu^2$.
After this new determination of the transverse potentials, we will see later
how we can determine $\mu^2$, using the transverse energy sum rule.\\
As noted in Ref.~\cite{bbs5}, if $p_z$ denotes the proton momentum, its energy can be approximated by $p_z + M^2/2p_z$, where $M$ is the proton mass.
Similarly the energy of a massless parton, with transverse momentum $k_T$ is, in the same approximation, $xp_z + k_T^2/2xp_z$. Therefore all
 involved parton distributions denoted $p_i(x,k_T^2)$ must satisfy the momentum sum rule
\begin{equation}
\sum_i \int_0^1dx \int x p_i(x,k^2_T) dk^2_T = 1~,
\label{msr}
\end{equation}
and also the transverse energy sum rule 
\begin{equation}
\sum_i \int_0^1dx \int p_i(x,k^2_T)\frac{k^2_T}{x}dk^2_T =
M^2 ~.
\label{esr}
\end{equation}
The contribution of Eq.~(\ref{nondifq}), for the quarks, to the sum rule
Eq.~(\ref{esr}) is given by:

\begin{equation}
\int\frac{k_T^2}{x} q^h (x , k_T^2) dx dk_T^2  =
\int_0^1 \frac{F(x)}{x^2 (\exp{\frac{x - X^h_{0q}}{\bar{x}}} + 1)} dx
\int_0^\infty \frac{k_T^2 d k_T^2}{\exp{(\frac{k_T^2}{ x \mu^2}- Y^h_{0q})} +
1}~,
\label{E2}
\end{equation}
and after the change of variable $\xi=k_T^2/x\mu^2$, we get
\begin{equation}
\mu^2 \int_0^1\frac{\mu^2 F(x) dx}{\exp{\frac{x - X^h_{0q}}{\bar{x}} + 1}}
\int_0^\infty\frac{
\xi d \xi}{\exp({\xi - Y^h_{0q}}) + 1} = \mu^2 I_1 \cdot I_2~,
\label{E3}
\end{equation}
where
\begin{equation}
I_1 = \int_0^1\frac{\mu^2 F(x) dx}{\exp{\frac{x - X^h_{0q}}{\bar{x}} +
1}}=\frac{P_{q^h}}{\ln{(1 + \exp{Y^h_{0q}})}}~,
\label{E4}
\end{equation}
$P_{q^h}$ is the number of parton of type $q^h$, and
\begin{equation}
I_2 = \int_0^\infty\frac{\xi d \xi}{\exp({\xi - Y^h_{0q}}) + 1}=
\frac{\pi^2}{6} +\frac{(Y^h_{0q})^2}{2} + \mbox{Li}_2(-\exp(-Y^h_{0q}))~.
\label{E5}
\end{equation}
Therefore in the limit $Y^h_{0q}=0$ the contribution of a parton of type $q^h$
is just $\mu^2 P_{q^h} \pi^2/(12 \ln{2})$, since $\mbox{Li}_2(-1)=-\pi^2/12$.\\

In a similar way for the contribution to the sum rule Eq.~(\ref{esr}), from the
non-diffractive part of the light antiquarks Eq.~(\ref{nondifqbar}), we get
\begin{equation}
\mu^2 \int_0^1\frac{\mu^2 \bar{F}(x) dx}{\exp{\frac{x + X^{-h}_{0q}}{\bar{x}} +
1}} \int_0^\infty\frac{
\xi d \xi}{\exp({\xi + Y^{-h}_{0q}}) + 1} = \mu^2 {\bar I}_1 \cdot {\bar I}_2~,
\label{E3bar}
\end{equation}
where
\begin{equation}
\bar {I}_1 = \int_0^1\frac{\mu^2 \bar {F}(x) dx}{\exp{\frac{x +
X^{-h}_{0q}}{\bar{x}} + 1}}=\frac{P_{\bar {q}^h}}{\ln{(1 +
\exp{(-Y^{-h}_{0q})})}}~,
\label{E4bar}
\end{equation}
$P_{\bar {q}^h}$ is the number of parton of type $\bar {q}^h$, and
\begin{equation}
\bar {I}_2 = \int_0^\infty\frac{\xi d \xi}{\exp({\xi + Y^{-h}_{0q}}) + 1}=
\frac{\pi^2}{6} +\frac{(Y^{-h}_{0q})^2}{2} + \mbox{Li}_2(-\exp(Y^{-h}_{0q}))~.
\label{E5bar}
\end{equation}

Finally we turn to the universal diffractive contribution to
quarks and antiquarks in Eqs.~(\ref{eq1},\ref{eq2}), namely
$xq^D(x,Q^2_0)=\tilde{A}x^{\tilde{b}}/[\exp(x/\bar{x}) + 1]$. Since
$\tilde{b}<0$ (see Eq.~(\ref{eq3})), the introduction
of the $k_T$ dependence cannot be done similarly to the non diffractive
contributions, because in the
energy sum rule Eq.~(\ref{esr}), it generates a singular behavior when $x \to
0$.
Therefore in order to avoid this difficulty, as in Ref.~\cite{bbs5}, we
modify our prescription by taking at the input energy scale
\begin{equation}
xq^D(x,k^2_T)=\frac{\tilde{A}x^{\tilde{b}-2}}{\mbox{ln}2\mu^2}
\frac{1}{[\exp(x/\bar{x}) + 1]}\frac{1}{[\exp(k^2_T/x^2\mu^2)+1]}~,
\label{diff}
\end{equation}
whose $k_T$ fall off is stronger, because $x\mu^2$ is now replaced by
$x^2\mu^2$. Note that this is properly
normalized to recover $xq^D(x,Q^2_0)$ after integration over $k_T^2$. We have
checked that $xq^D(x,k^2_T)$ gives
a negligible contribution to Eq.~(\ref{esr}), as expected (See Appendix).
Concerning the gluon, since it is parametrized by a quasi
Bose-Einstein function, one
has to introduce a non-zero potential $Y_G$, in contrast with the QCD
equilibrium conditions, to avoid the singular behavior of
$\mbox{Li}_1(\exp[-Y_G/\bar{x}])$, when
$Y_G=0$. The value of $Y_G$ is not constrained, but by taking a
very small $Y_G$, it does not affect the energy sum rule (See Appendix).\\
Clearly these regularisation procedures to the diffractive contribution and to
the gluon are not fully satisfactory, but we will discuss them again in our
concluding remarks (See Section 5).
By summing up all contributions to the energy sum rule, one finally gets the
value of $\mu^2$, namely $\mu^2=0.198 \mbox{GeV}^2$.

\subsection{The Melosh-Wigner transformation}
So far in all our quark or antiquark TMD distributions (see
Eqs.~(\ref{nondifq},\ref{nondifqbar})), the label "`$h$"' stands for the
helicity along the
longitudinal momentum and not along the direction of the momentum, as normally
defined for a genuine helicity. The basic effect of a transverse momentum $k_T
\neq 0$ is the Melosh-Wigner rotation \cite{mel-wig,bucc}, which mixes the
components $q^{\pm}$ in the following way
\begin{equation}
q^{+'}= \cos^2\theta ~q^+ + \sin^2\theta ~q^- ~~~~\mbox{and}~~~q^{-'}=
\cos^2\theta ~q^- + \sin^2\theta ~q^+,
\label{mel1}
\end{equation}
where $2\theta=\mbox{Arctg}(\kappa k_T/xM)$, $M$ is the proton mass and
$\kappa$ is a dimensionless parameter.\\
Consequently $q = q^+ + q^-$ remains unchanged $q'=q$, whereas we have
\begin{equation}
\Delta q'= (\mbox{cos}^2\theta - \mbox{sin}^2\theta) ~\Delta q =
\mbox{cos}2\theta ~\Delta q = \mbox{cos~Arctg}(\kappa k_T/xM)~\Delta q~.
\label{mel2}
\end{equation}
So we finally get
\begin{equation}
\Delta q'= \frac{1}{\sqrt{1 + (\kappa k_T/xM)^2}}~\Delta q~.
\label{mel3}
\end{equation}
The effect of the Melosh-Wigner transformation on the double longitudinal-spin
asymmetry will be discussed in Section 4.

 \subsection{The TMD distributions in the relativistic covariant approach}
\label{tmdcov}
Covariant parton models have been widely discussed in the literature, but in some recent papers \cite{zav,estz}, an analysis based
on the requirements of symmetry, for the parton motion in the nucleon rest frame, leads to
a different method to generate the TMD of a given $x$-distribution. By using some input
unpolarized distribution $f(x)$, one can calculate the corresponding TMD
distribution $f(x,k^2_T)$, by means of its
derivative, according to the following rule
\begin{equation}
f(x,k^2_T)=- \frac{1}{\pi M^2}\frac{d}{d\xi}(f(\xi)/\xi)~,
\label{cov1}
\end{equation}
where the variable $\xi$ is defined as $\xi = x(1 + k_T^2/x^2M^2)$, $M$ being
the proton mass.\\
This method has been generalized for helicity distributions 
$\Delta f(x)$ and in this case we have for the corresponding TMD distribution
$\Delta f(x,k^2_T)$
\begin{equation}
\Delta f(x,k^2_T)= \frac{2x - \xi}{\pi M^2 \xi^3}[3\Delta f(\xi) + 2
\int_\xi^1\frac{\Delta f(y)}{y}dy -\xi \frac{d}{d\xi}\Delta f(\xi)]~.
\label{cov2}
\end{equation}
It is interesting to recall that in Ref.~\cite{alm}, it was demonstrated that for the
 TMD PDF a factorized form $f_1(x)f_2(k_T^2)$ is in contradiction with Lorentz structure, at least for zero strong interaction coupling $g=0$. Using a rather different approach, they obtain results identical to the above ones.\\
We will show and discuss later the results one obtains from these formulas,
 using as input the
$x$-dependent statistical PDF in Eqs.~(1,2). In this approach as well as in
ours, one gets distributions function of $x$ and $k_T^2/x$.

\section{Cross section and spin asymmetry of pion production
in polarized SIDIS}
Following Ref.~\cite{kot}, we consider the polarized SIDIS, $\ell \, N \to \ell
\, H \, X$ in the
simple quark-parton model, with unintegrated parton distributions.
According to  the standard notations for DIS variables, $\ell$ and $\ell'$
are, respectively, the four-momenta of the initial and the final
state leptons, $q = \ell - \ell'$ is the exchanged virtual photon
momentum, $P$ is the target nucleon momentum, $P_H$ is the final hadron
momentum, $Q^2=-q^2$,
 $x=Q^2/2P\cdot q$, $y=P\cdot q/P\cdot \ell$, $z=P\cdot P_H/P\cdot
q$, $Q^2 = xy(s - M^2)$ and $s = (\ell + P)^2$. We work in a frame with
the $z$-axis along the virtual photon momentum direction and the
$x$-axis in the lepton scattering plane, with positive direction
chosen along the lepton transverse momentum. The produced hadron has
transverse momentum $p_{T}$ (For further details see Ref. \cite{kot}).\\
Keeping only twist-two contributions and terms up to $\mathcal{O}(M/Q)$, the
cross section for SIDIS of longitudinally polarized leptons off a
longitudinally polarized target can be written as:

\begin{equation}\label{sig}
\frac{d^5\sigma^{\begin{array}{c}\hspace*{-0.1cm}\to\vspace*{-0.25cm}\\
\hspace*{-0.1cm}\Leftarrow\end{array}}}{dx \, dy \, dz \, d^2p_{T}} =
\frac{2 \alpha^2}{x y^2 s}\, \left\{ {\cal H}_{1} + \lambda \, S_L{\cal
H}_{2}\right\}~,
\end{equation}
where the arrows indicate the direction of the lepton
($\rightarrow$) and target nucleon ($\Leftarrow$) polarizations,
with respect to the lepton momentum; $\lambda$, and $S_L$ are
the magnitudes of the longitudinal beam polarization and
the longitudinal target polarization, respectively.

The two terms have the following simple partonic expressions
\begin{equation}\label{hf1p}
{\cal H}_1(p_{T}) = \sum_q e_q^2 \int d^2 k_T q(x,k_T) \, \pi
y^2 \, \frac{\hat s^2 + \hat u^2} {Q^4} \, D_q^h(z, q_T),
\end{equation}
\begin{equation}\label{hg1lp}
{\cal H}_2(p_{T}) = \sum_q e_q^2 \int d^2 k_T\Delta q'(x, k_T) \,
\pi y^2 \, \frac{\hat s^2 - \hat u^2} {Q^4} \, D_q^h(z, q_T),
\end{equation}
where $p_{T}= q_T + zk_T$ and $q_T$ is the intrinsic transverse momentum of the
hadron $H$ with
respect to the fragmenting quark direction. Here $\hat s$, $\hat t$ and $\hat
u$ are the Mandelstam variables for the subprocess $\ell q \rightarrow \ell q$.
Note that in Eq.~(\ref{hg1lp}) above, we have used Eq.~(\ref{mel3}), which
takes into account the effect of the Melosh-Wigner rotation.\\
The first two contributions, Eqs. (\ref{hf1p}) and (\ref{hg1lp}), give,
respectively, the unpolarized cross section and the numerator of the double
longitudinal-spin asymmetry $A_1$
\begin{equation}
\label{2terms}
\frac{d^5\sigma}{dx \, dy \, dz \, d^2 p_{T}} = \frac{2 \alpha^2}
{x \, y^2s} \> {\cal H}_1 \quad\quad
\frac{d^5\sigma^{++}} {dx \, dy \, dz \, d^2 p_{T}} -
\frac{d^5\sigma^{+-}} {dx \, dy \, dz \, d^2 p_{T}} =
\frac{4 \alpha^2} {x \, y^2s} \> {\cal H}_2 \>,
\end{equation}
where $+,-$ stand for helicity states.  So we simply have $A_1 = 2{\cal
H}_2/{\cal H}_1$.

The integrals in Eqs. (\ref{hf1p},\ref{hg1lp}) involve the following TMD
fragmentation function \cite{kretzer}
\begin{equation}
D_q^h(z, q_T)= D_q^h(z) \, \frac{1}{\pi \mu_D^2}\,
\exp\left( -\frac{q_T^2}{\mu_D^2} \right)\label{dfffg2},
\end{equation}
which is the standard factorized Gaussian model, since we have not yet generalized
our statistical approach to the TMD fragmentation functions.

\section{Results and discussion}
\begin{figure}[htbp]
\begin{center}
  \begin{minipage}{6.5cm}
  \epsfig{figure=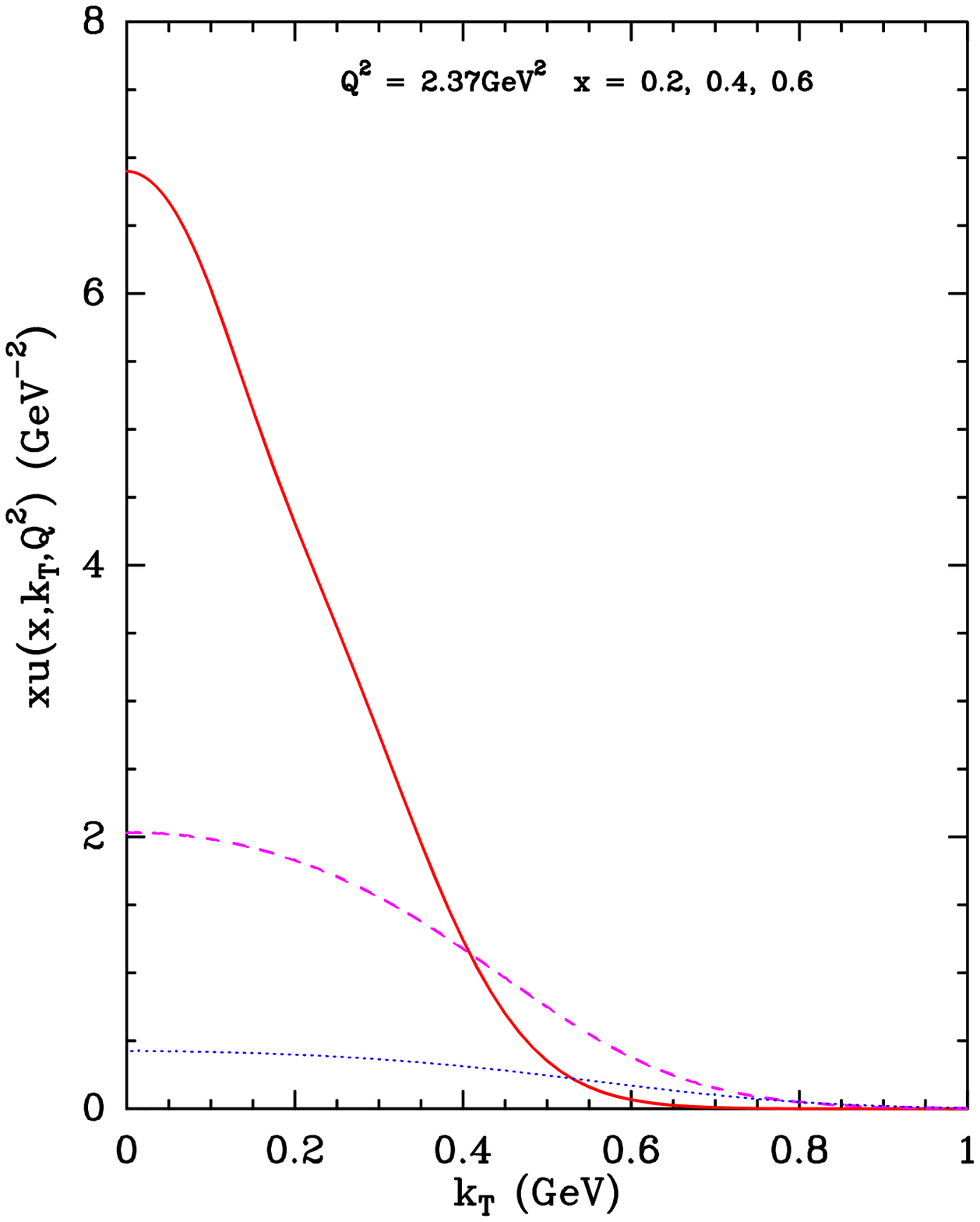,width=8.5cm}
  \end{minipage}
    \begin{minipage}{6.5cm}
  \epsfig{figure=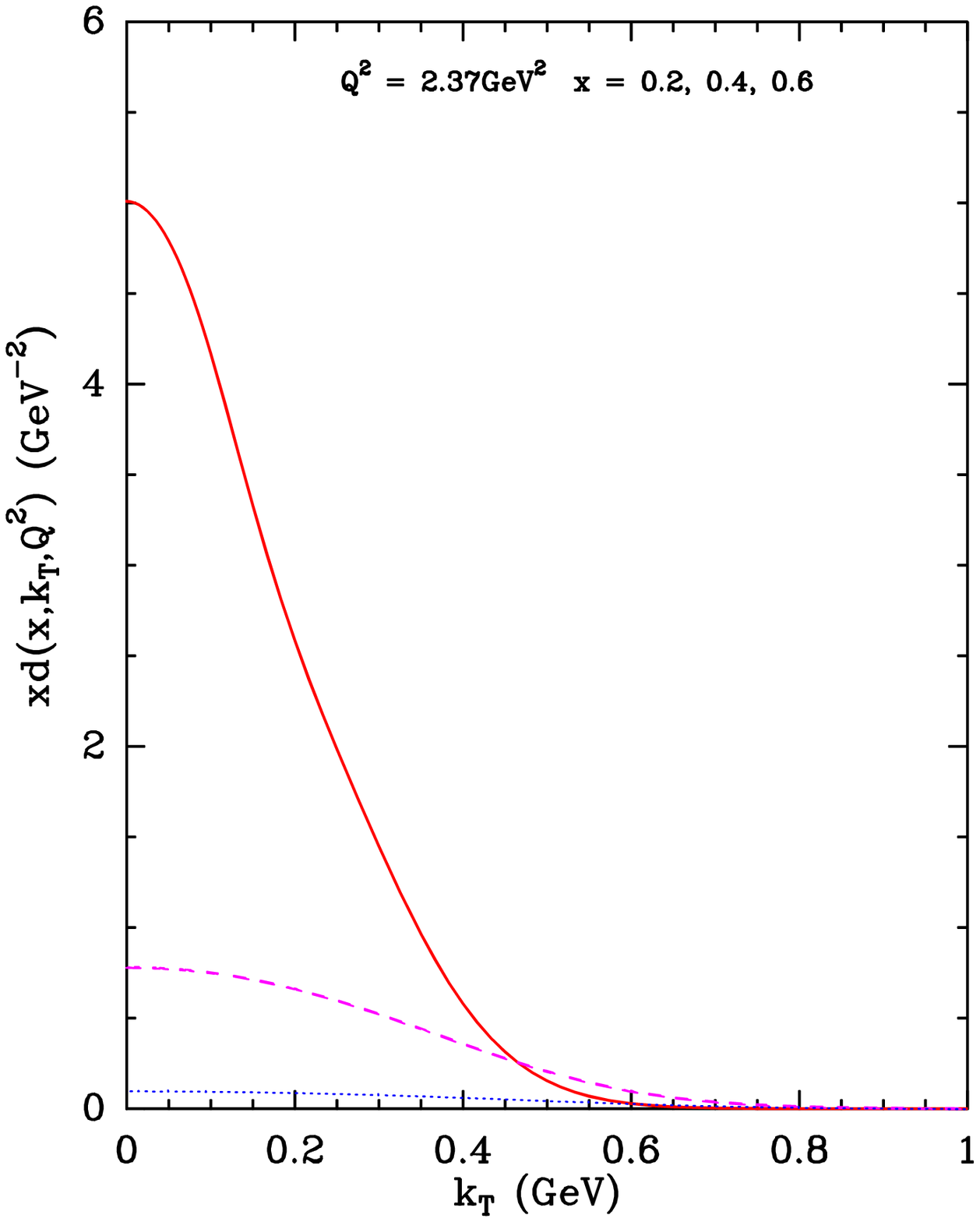,width=8.5cm}
    \end{minipage}
\end{center}
  \vspace*{-10mm}
\caption{
The statistical distributions $xu(x,k_T,Q^2)$ ($\it left$) and $xd(x,k_T,Q^2)$
($\it right$), calculated at $Q^{2}=2.37~\mbox{GeV}^2$, versus $k_T$, for
different $x$ values: solid line $x=0.2$, dashed line $x=0.4$, dotted line
$x=0.6$}
\label{bbsud}
\vspace*{-1.5ex}
\end{figure}

\begin{figure}[htbp]
\begin{center}
  \begin{minipage}{6.5cm}
  \epsfig{figure=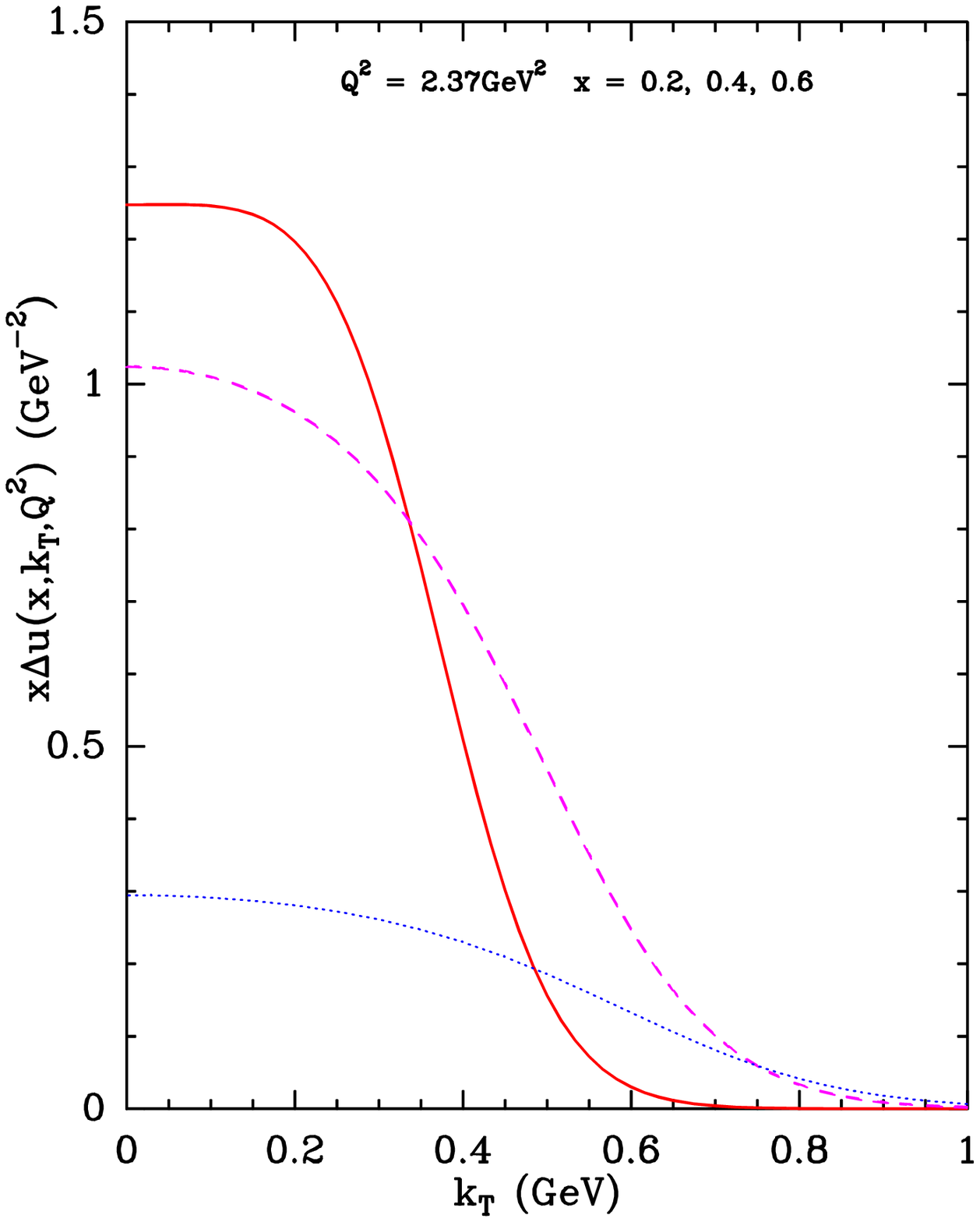,width=8.2cm}
  \end{minipage}
    \begin{minipage}{6.5cm}
  \epsfig{figure=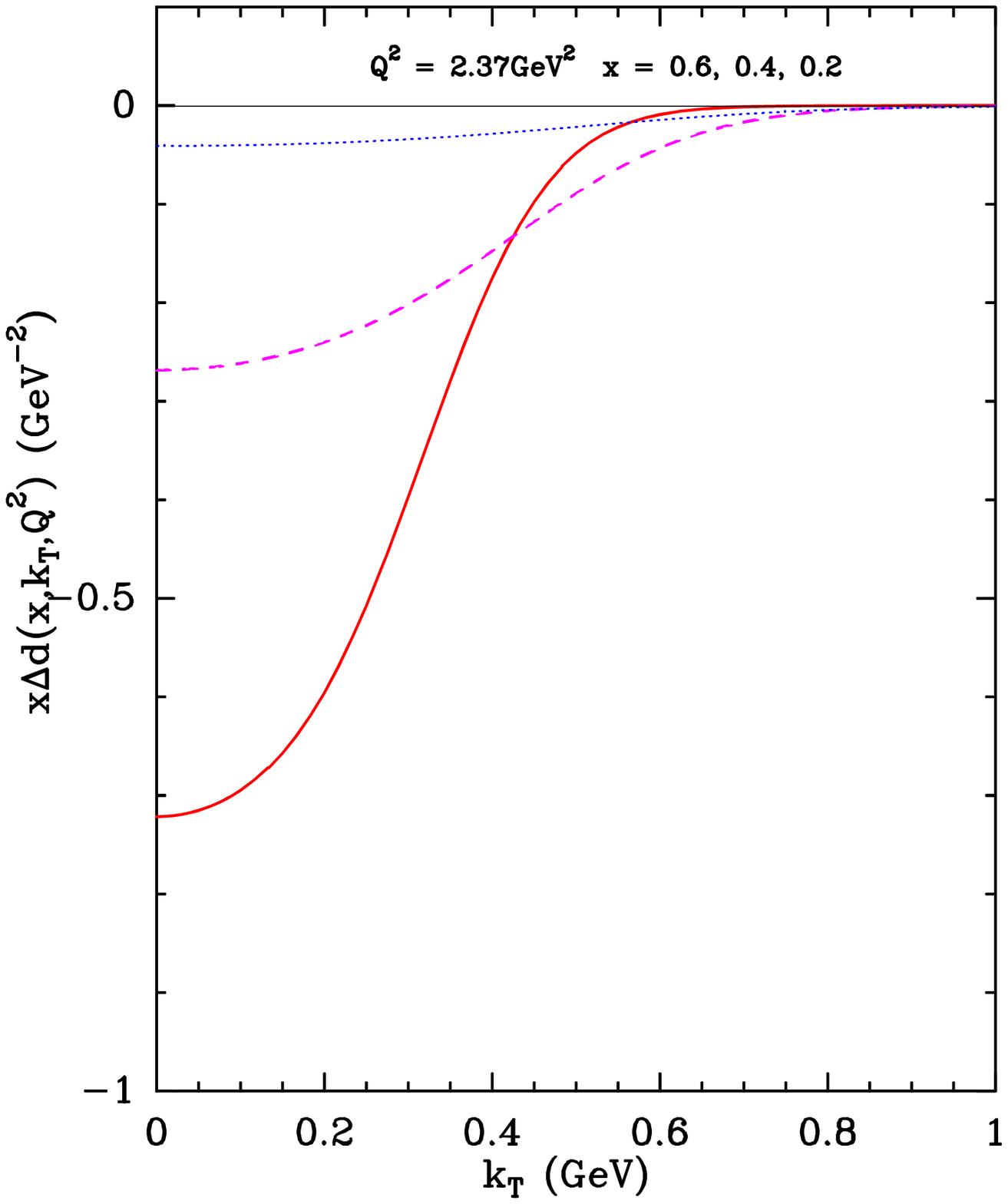,width=8.2cm}
    \end{minipage}
\end{center}
  \vspace*{-10mm}
\caption{
The statistical distributions $x\Delta u(x,k_T,Q^2)$ ($\it left$) and $x\Delta d(x,k_T,Q^2)$
($\it right$), calculated at $Q^{2}=2.37~\mbox{GeV}^2$, versus $k_T$, for
different $x$ values: solid line $x=0.2$, dashed line $x=0.4$, dotted line
$x=0.6$}
\label{bbsdelud}
\vspace*{-1.5ex}
\end{figure}

\begin{figure}[htbp]
\begin{center}
  \begin{minipage}{6.5cm}
  \epsfig{figure=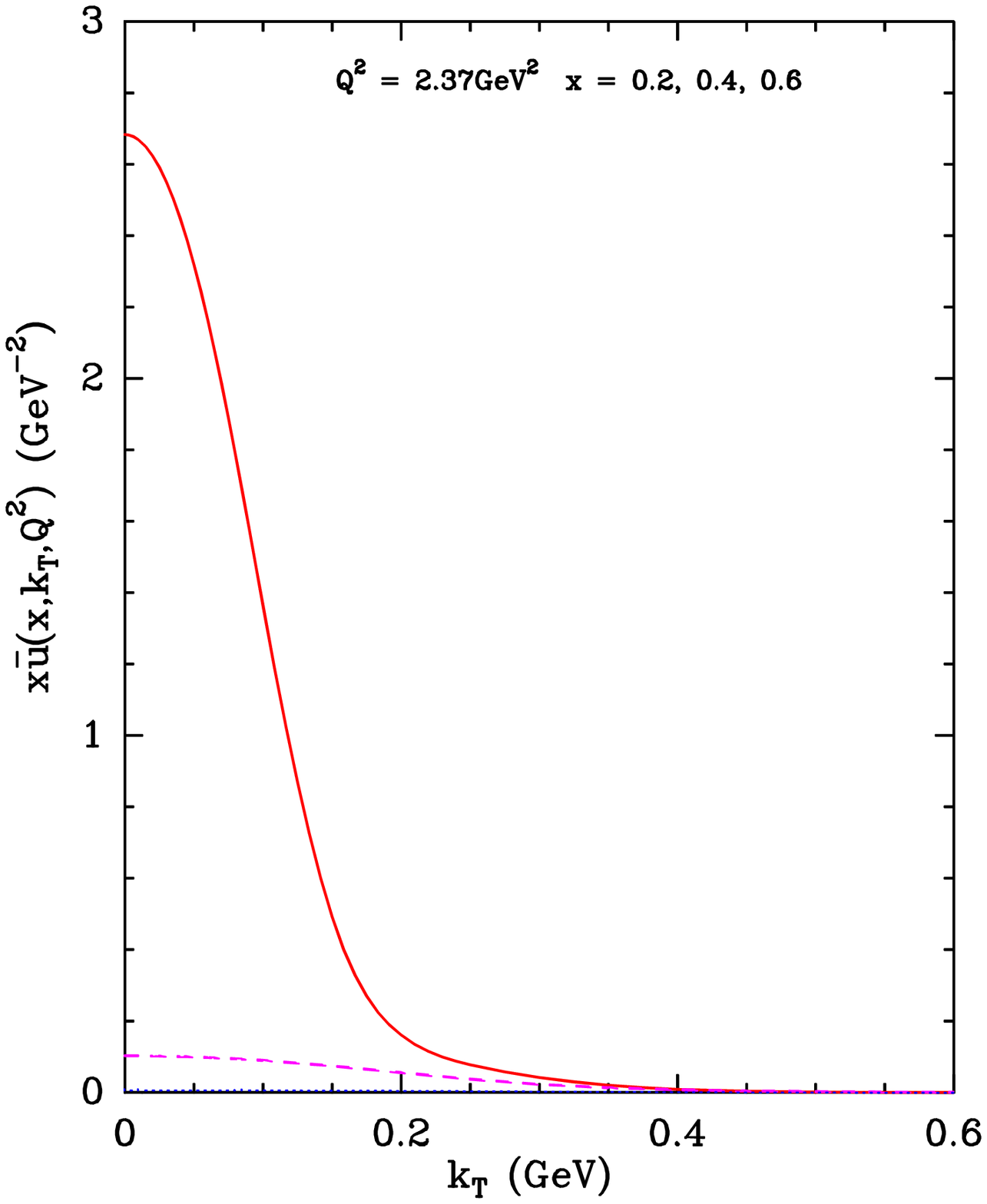,width=8.5cm}
  \end{minipage}
    \begin{minipage}{6.5cm}
  \epsfig{figure=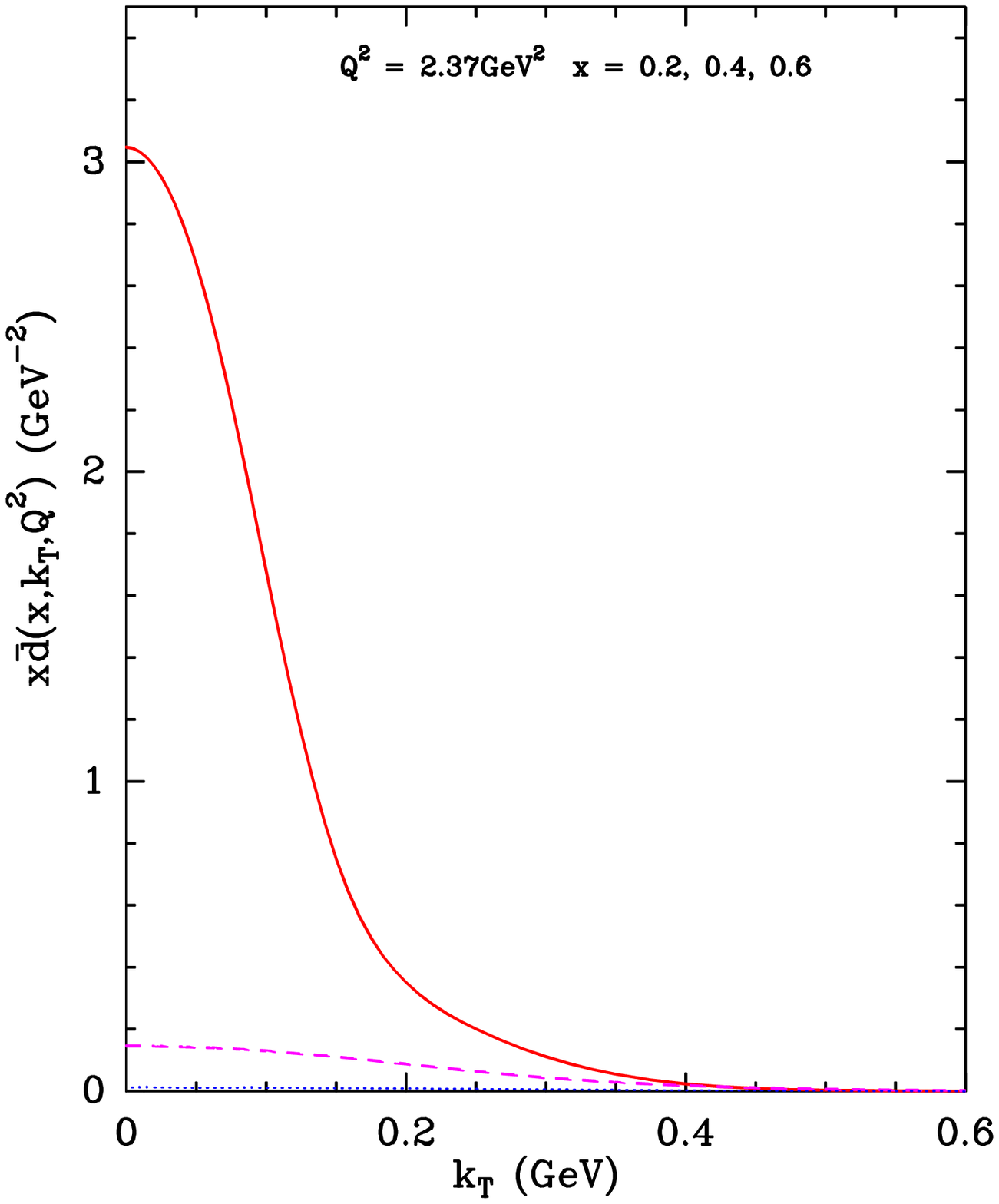,width=8.5cm}
    \end{minipage}
\end{center}
  \vspace*{-10mm}
\caption{
The statistical distributions $x\bar {u}(x,k_T,Q^2)$ ($\it left$) and $x\bar {d}(x,k_T,Q^2)$
($\it right$), calculated at $Q^{2}=2.37~\mbox{GeV}^2$, versus $k_T$, for
different $x$ values: solid line $x=0.2$, dashed line $x=0.4$, dotted line
$x=0.6$}
\label{bbsbarud}
\vspace*{-1.5ex}
\end{figure}

\begin{figure}[htbp]
\begin{center}
  \begin{minipage}{6.5cm}
  \epsfig{figure=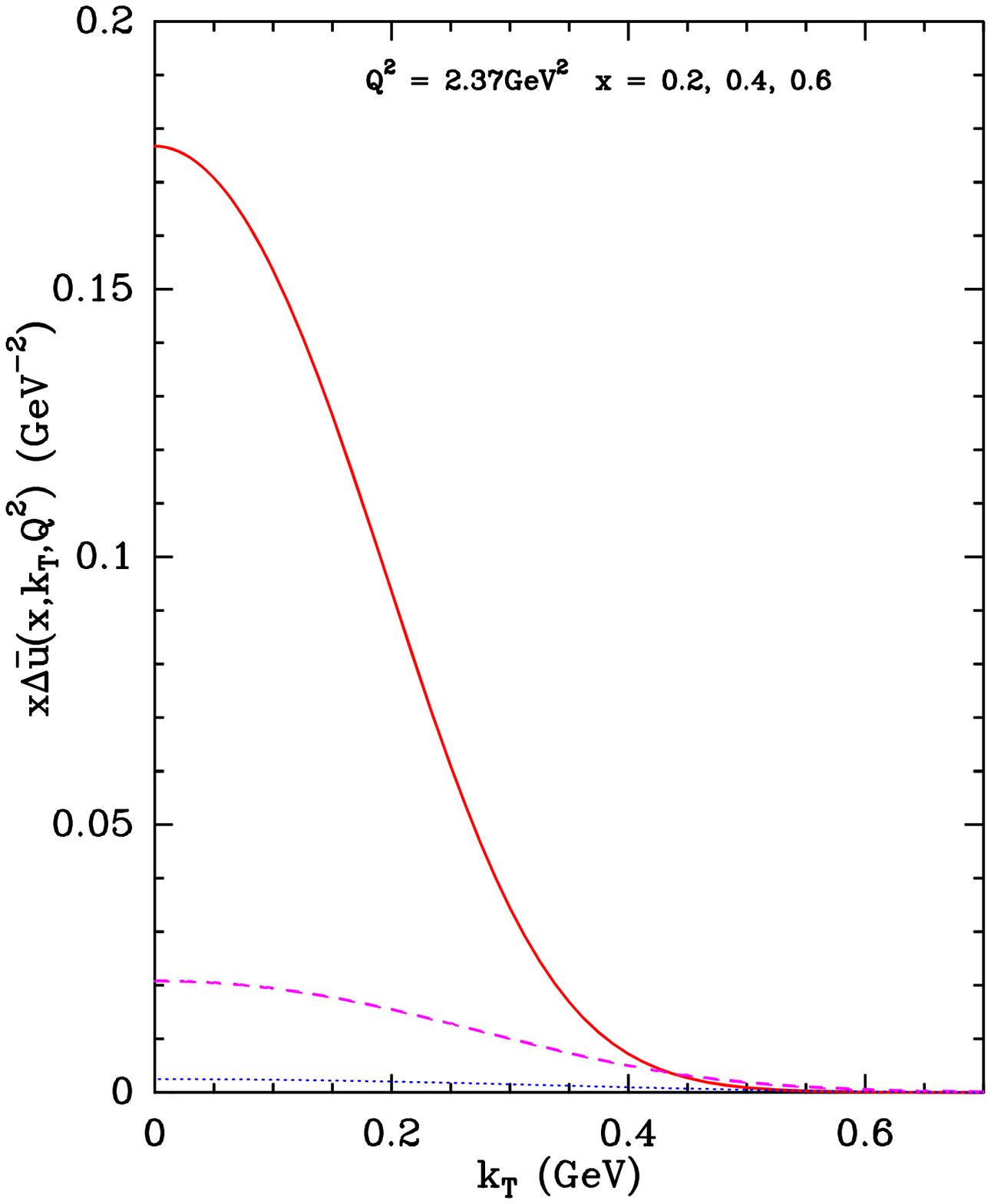,width=8.2cm}
  \end{minipage}
    \begin{minipage}{6.5cm}
  \epsfig{figure=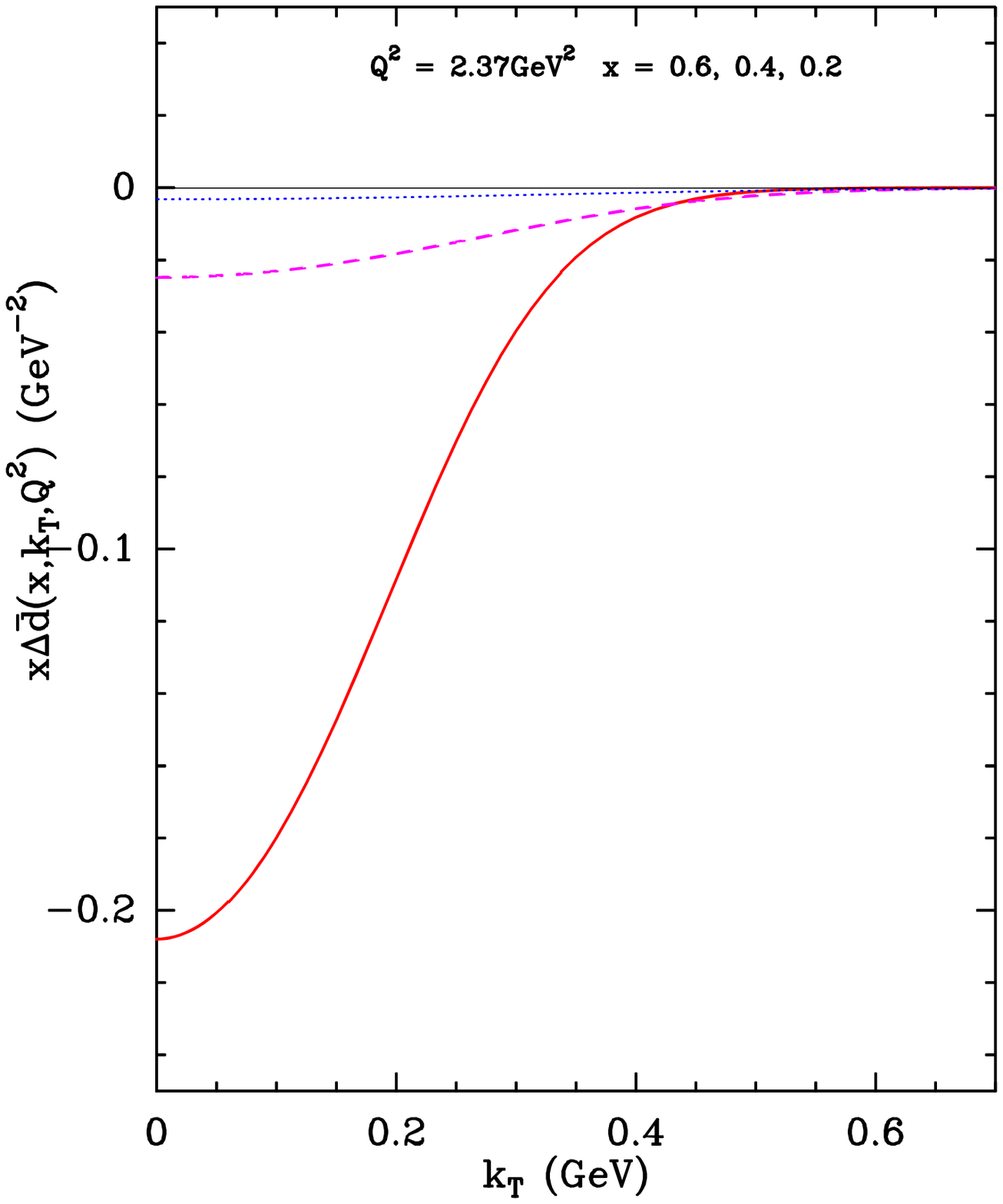,width=8.2cm}
    \end{minipage}
\end{center}
  \vspace*{-10mm}
\caption{
The statistical distributions $x\Delta \bar{u}(x,k_T,Q^2)$ ($\it left$) and $x\Delta \bar{d}(x,k_T,Q^2)$
($\it right$), calculated at $Q^{2}=2.37~\mbox{GeV}^2$, versus $k_T$, for
different $x$ values: solid line $x=0.2$, dashed line $x=0.4$, dotted line
$x=0.6$}
\label{bbsbardelud}
\vspace*{-1.5ex}
\end{figure}

\begin{figure}[htbp]
\begin{center}
  \begin{minipage}{6.5cm}
  \epsfig{figure=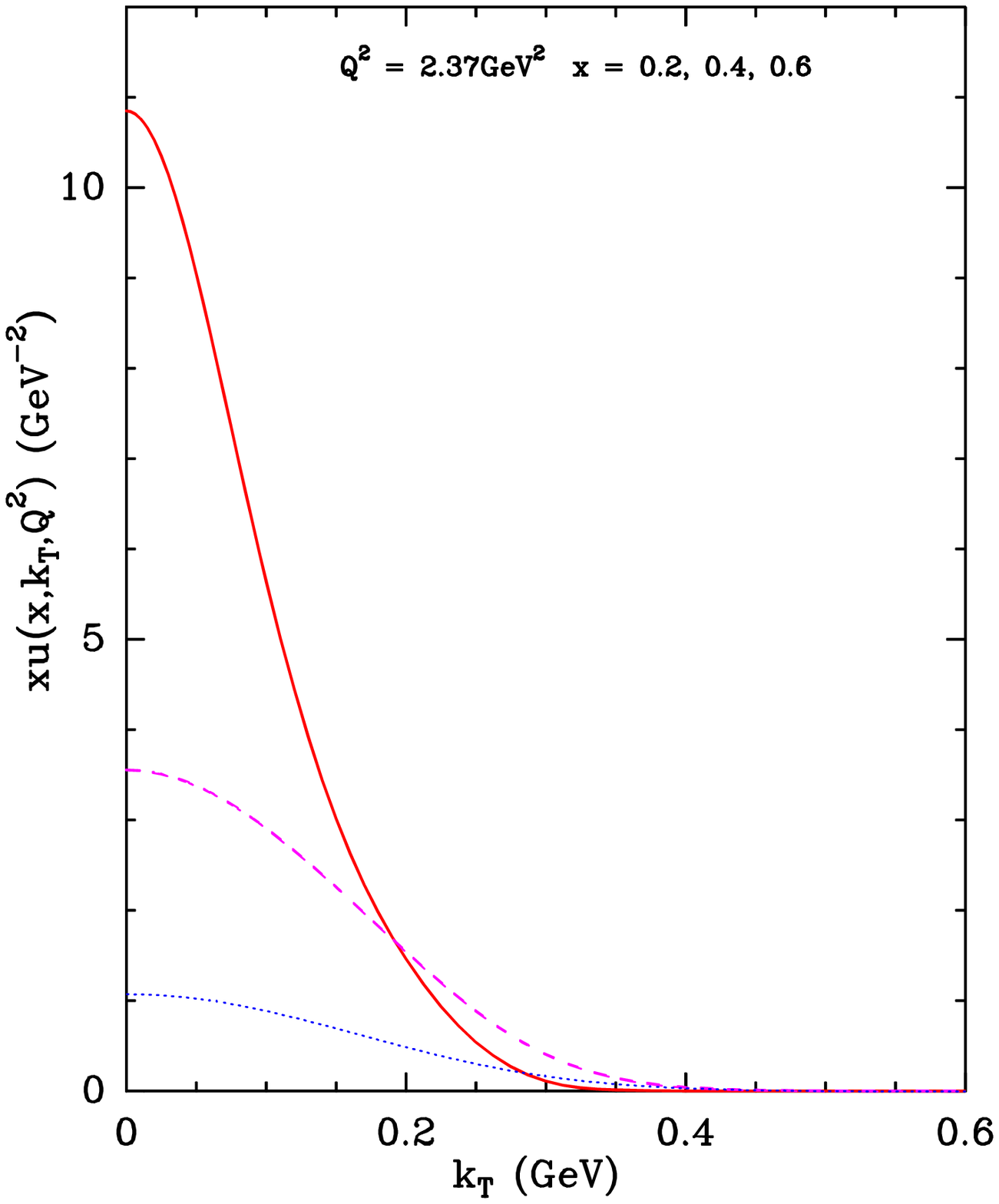,width=8.5cm}
  \end{minipage}
    \begin{minipage}{6.5cm}
  \epsfig{figure=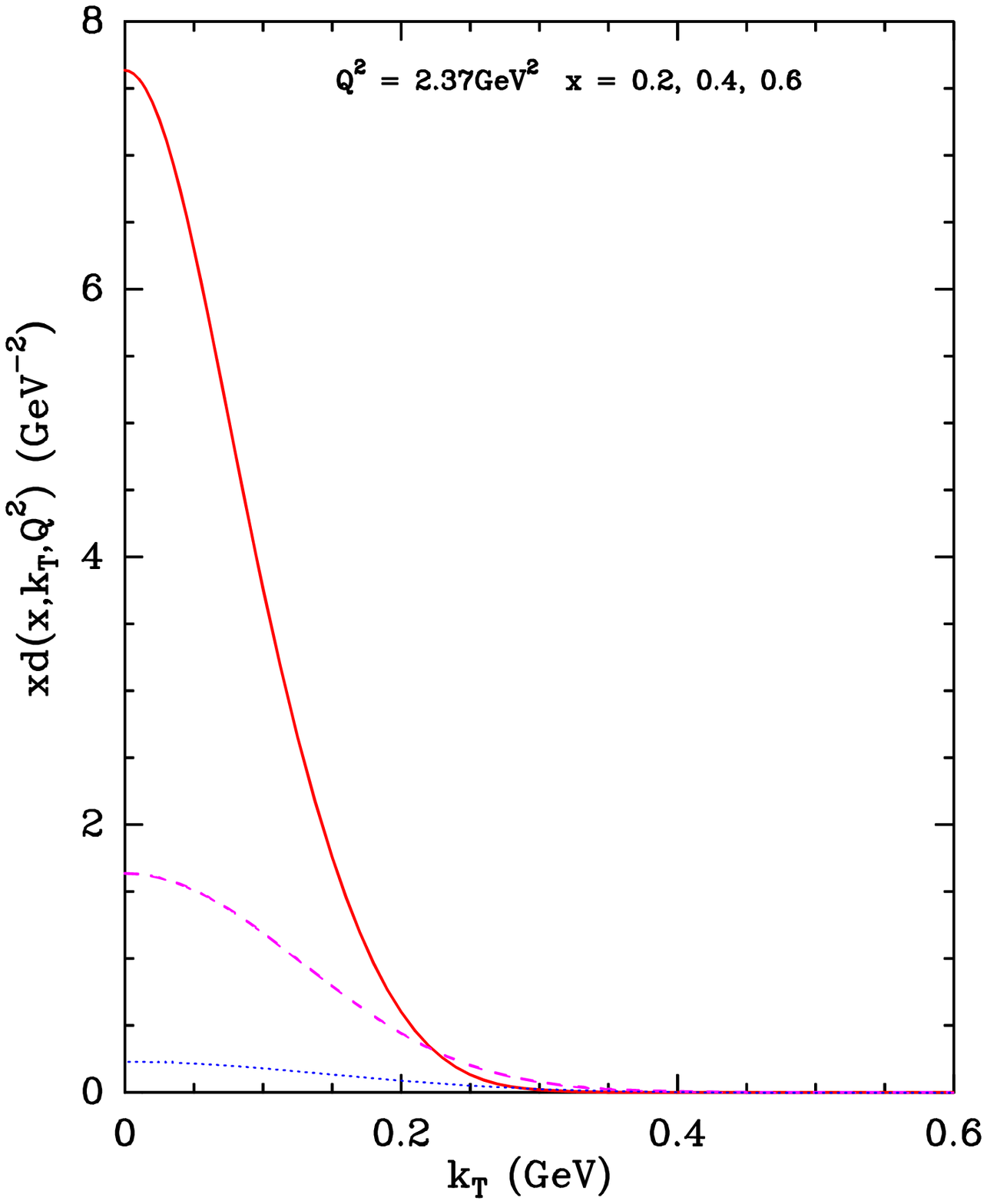,width=8.5cm}
    \end{minipage}
\end{center}
  \vspace*{-10mm}
\caption{
The relativistic covariant distributions $xu(x,k_T,Q^2)$ ($\it left$) and $xd(x,k_T,Q^2)$
($\it right$), calculated at $Q^{2}=2.37~\mbox{GeV}^2$, versus $k_T$, for
different $x$ values: solid line $x=0.2$, dashed line $x=0.4$, dotted line
$x=0.6$}
\label{zavud}
\vspace*{-1.5ex}
\end{figure}

\begin{figure}[htbp]
\begin{center}
  \begin{minipage}{6.5cm}
  \epsfig{figure=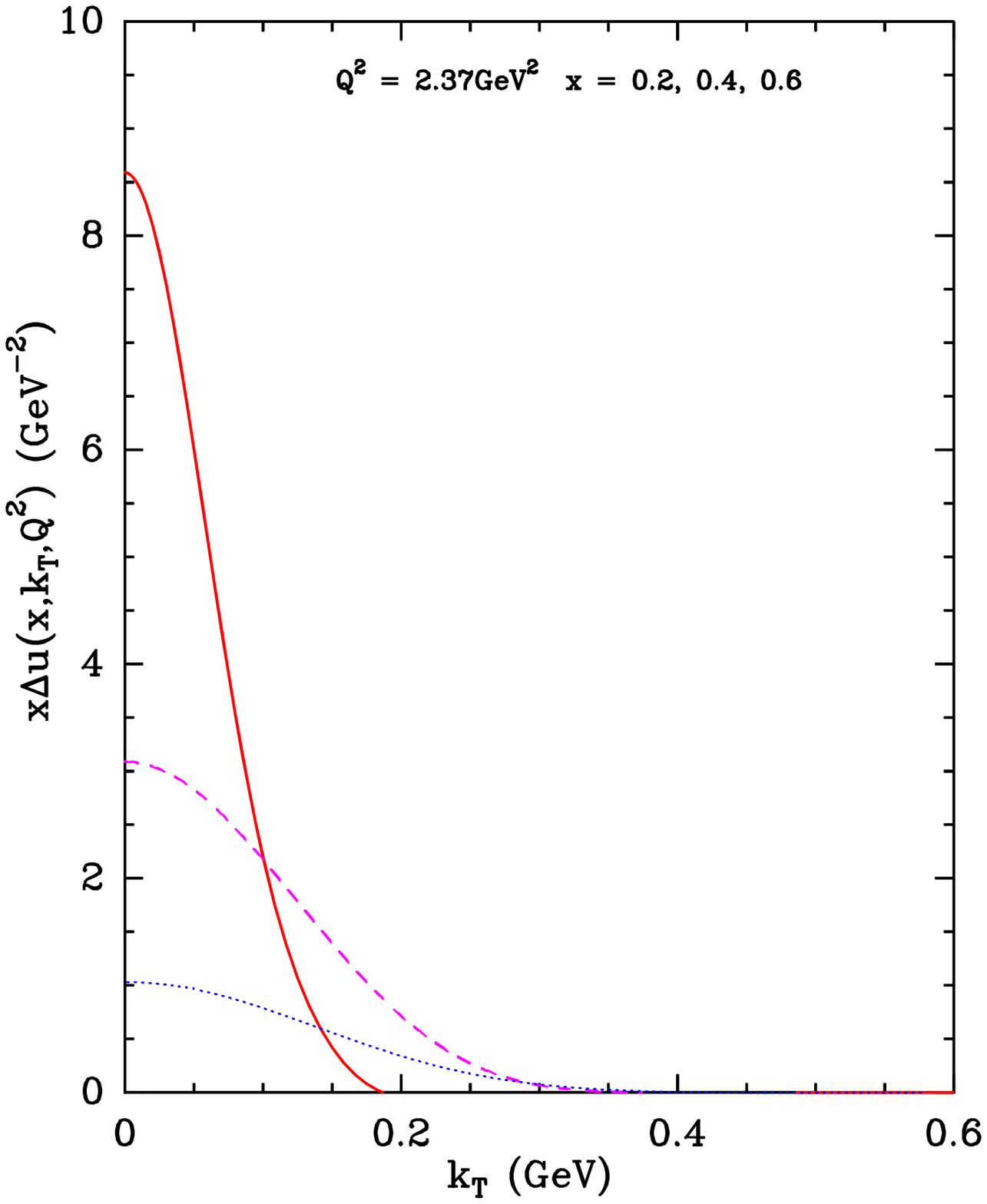,width=8.2cm}
  \end{minipage}
    \begin{minipage}{6.5cm}
  \epsfig{figure=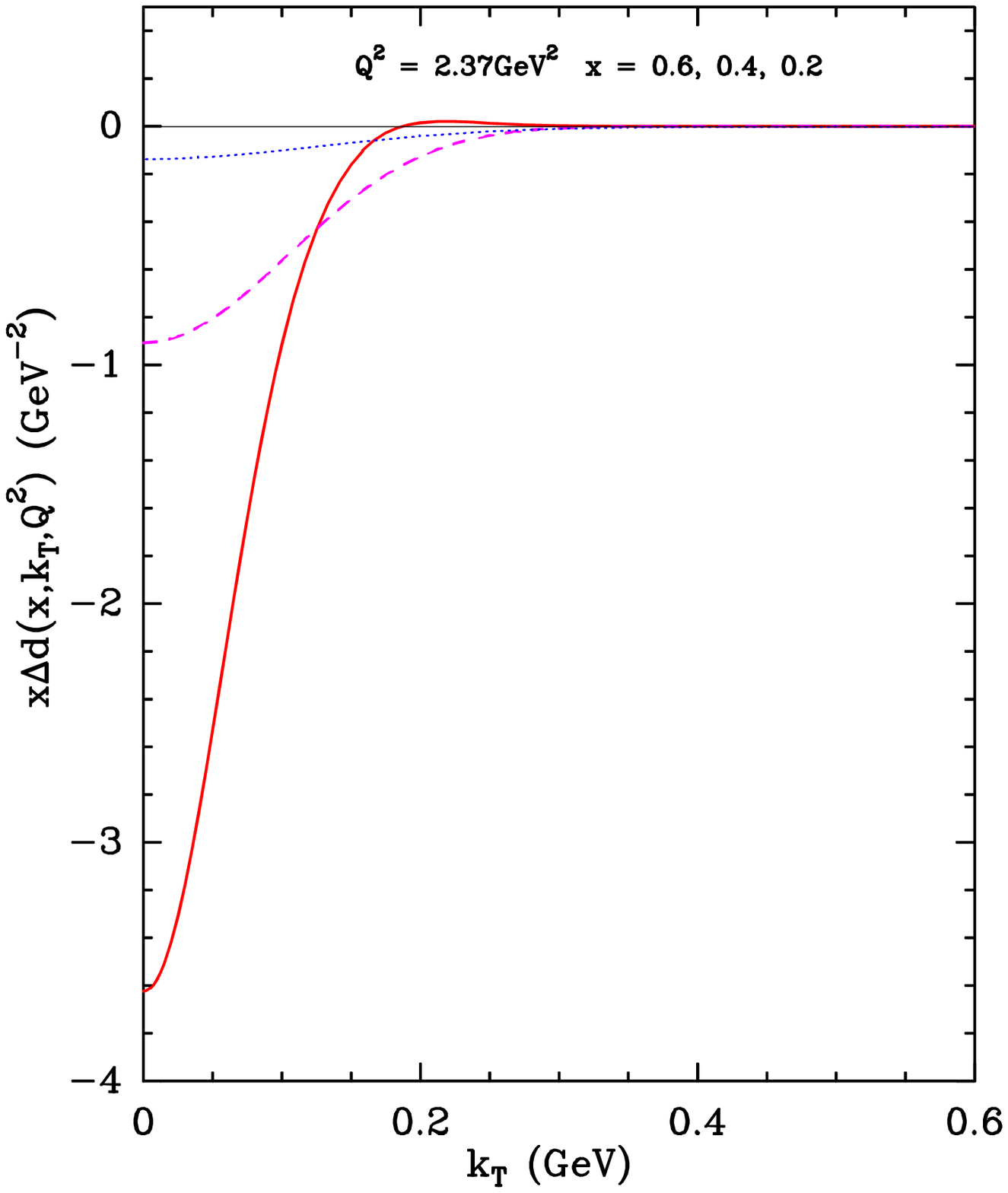,width=8.2cm}
    \end{minipage}
\end{center}
  \vspace*{-10mm}
\caption{
The relativistic covariant distributions $x\Delta u(x,k_T,Q^2)$ ($\it left$) and $x\Delta d(x,k_T,Q^2)$
($\it right$), calculated at $Q^{2}=2.37~\mbox{GeV}^2$, versus $k_T$, for
different $x$ values: solid line $x=0.2$, dashed line $x=0.4$, dotted line
$x=0.6$}
\label{zavdelud}
\vspace*{-1.5ex}
\end{figure}

\begin{figure}[htbp]
\begin{center}
  \begin{minipage}{6.5cm}
  \epsfig{figure=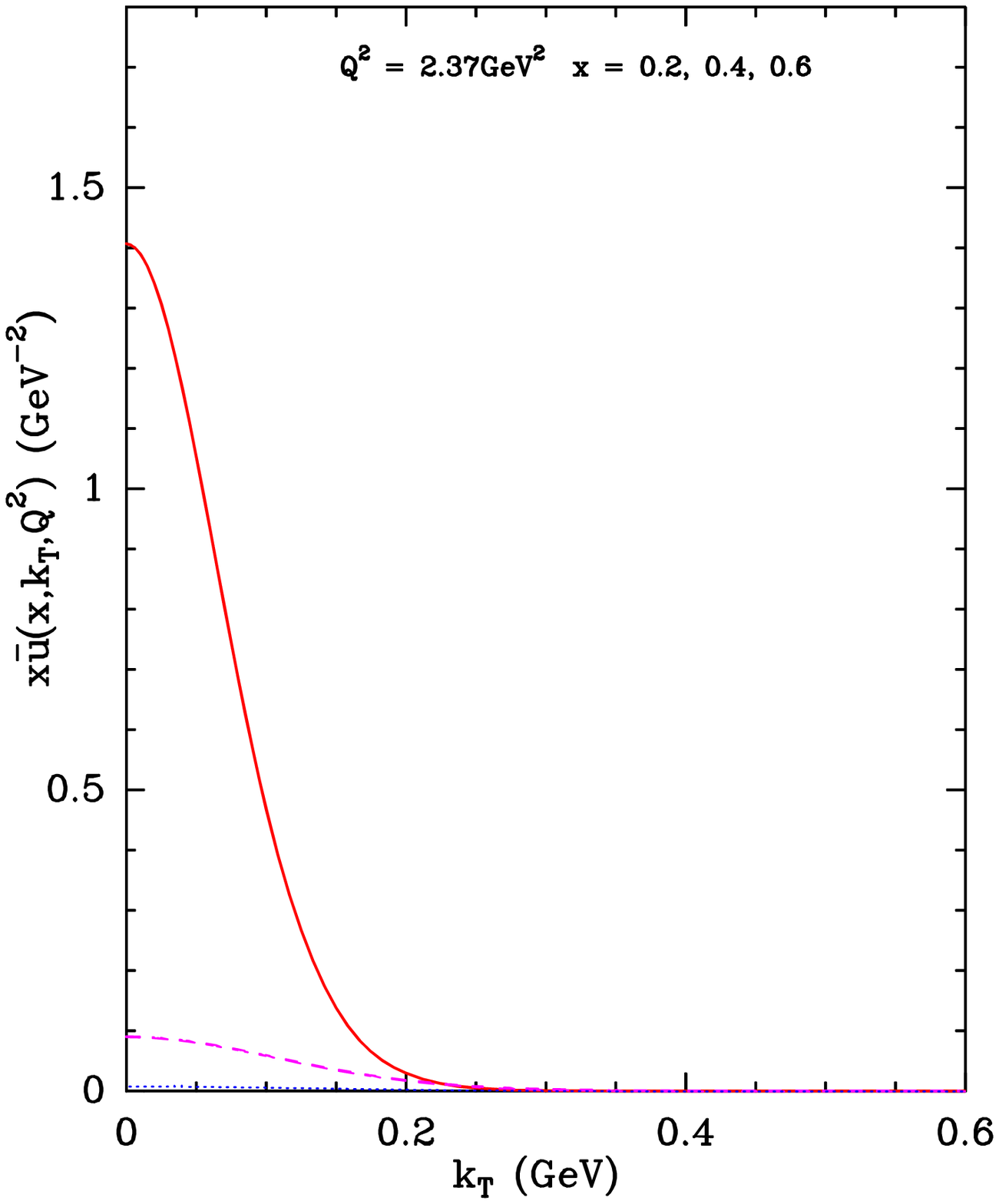,width=8.5cm}
  \end{minipage}
    \begin{minipage}{6.5cm}
  \epsfig{figure=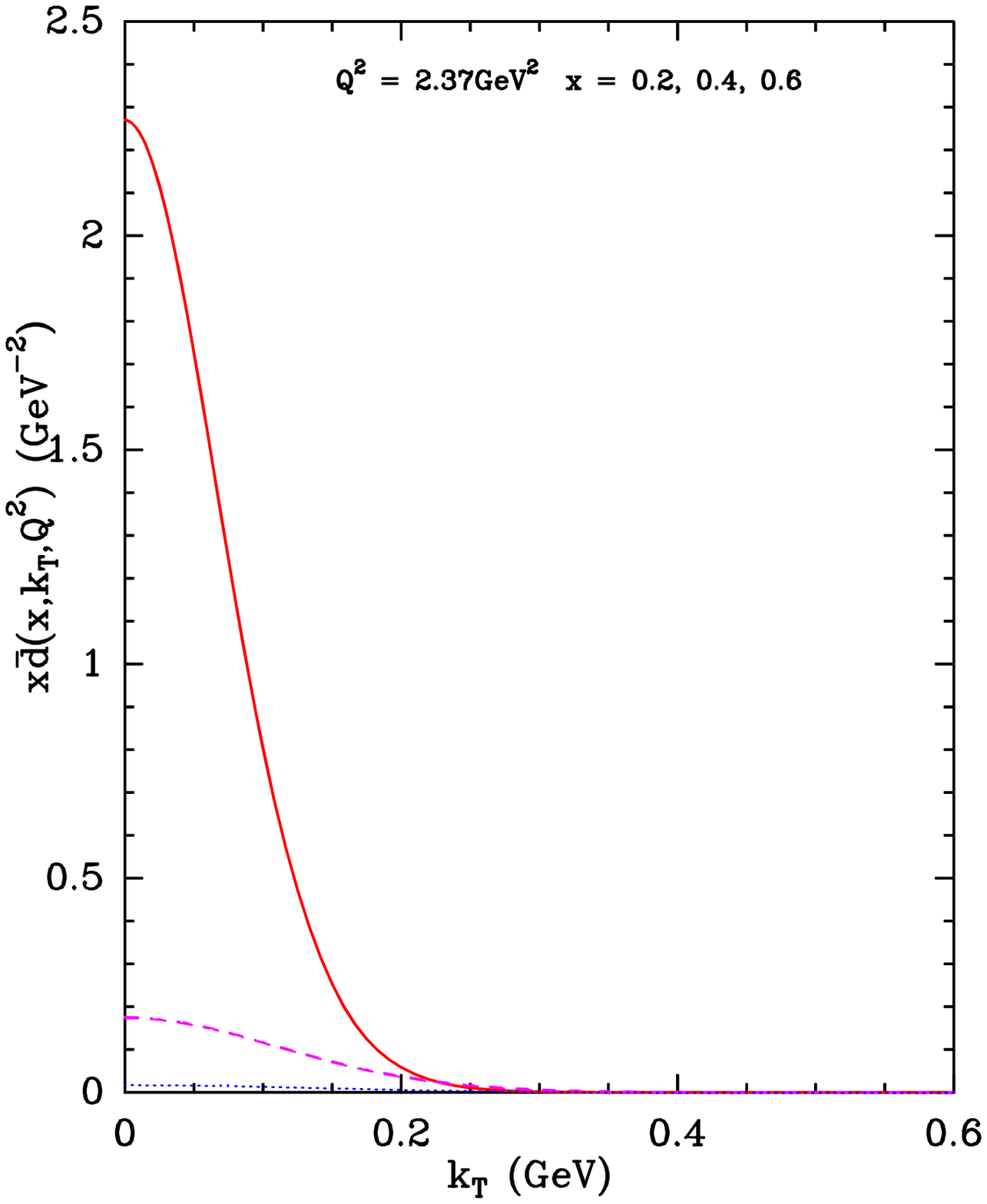,width=8.5cm}
    \end{minipage}
\end{center}
  \vspace*{-10mm}
\caption{
The relativistic covariant distributions $x\bar u(x,k_T,Q^2)$ ($\it left$) and
$x\bar d(x,k_T,Q^2)$ ($\it right$), calculated at $Q^{2}=2.37~\mbox{GeV}^2$, 
versus $k_T$, for different $x$ values: solid line $x=0.2$, dashed line $x=0.4$,
dotted line $x=0.6$.}
\label{zavbarud}
\vspace*{-1.5ex}
\end{figure}

\begin{figure}[htbp]
\begin{center}
  \begin{minipage}{6.5cm}
  \epsfig{figure=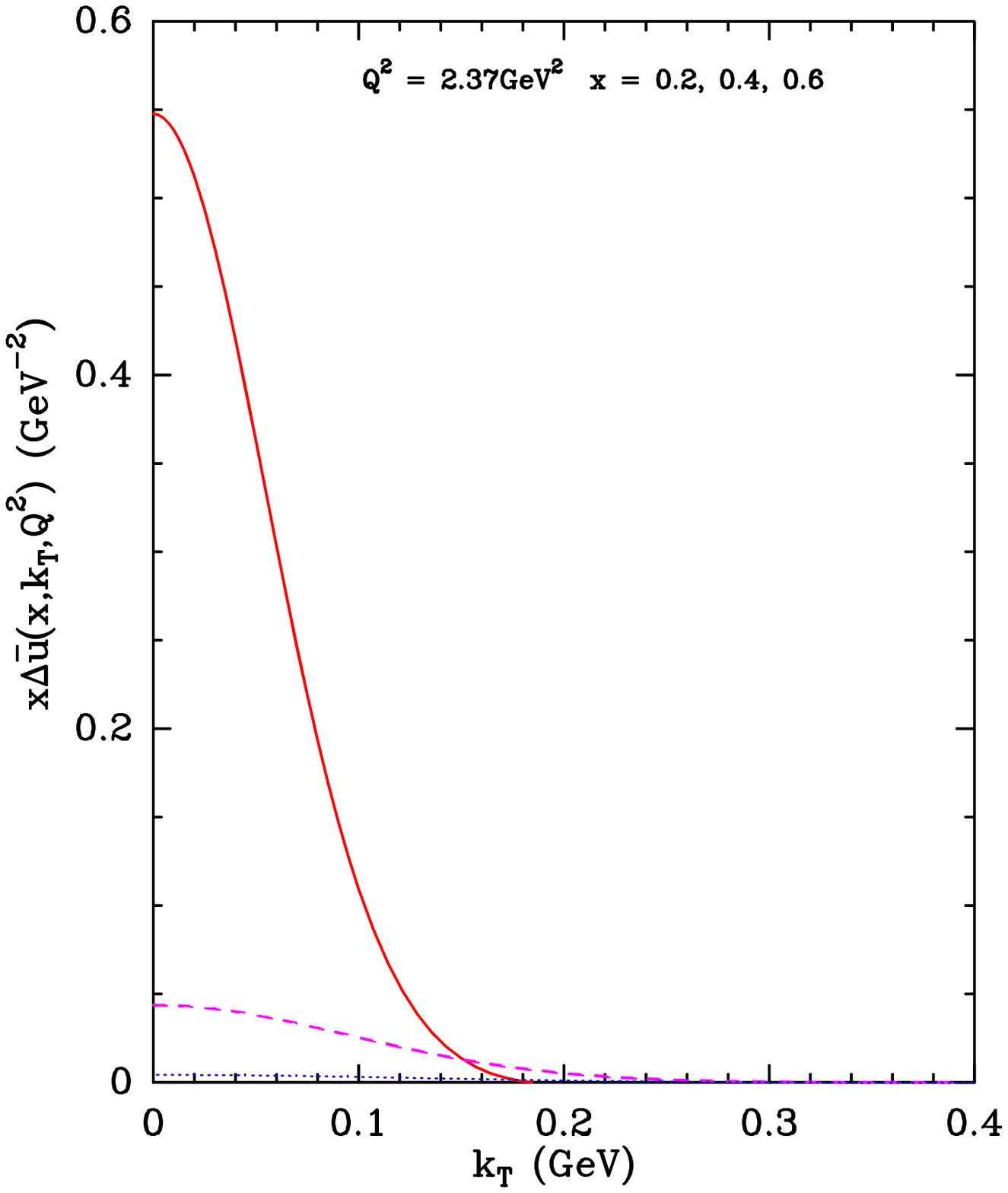,width=8.2cm}
  \end{minipage}
    \begin{minipage}{6.5cm}
  \epsfig{figure=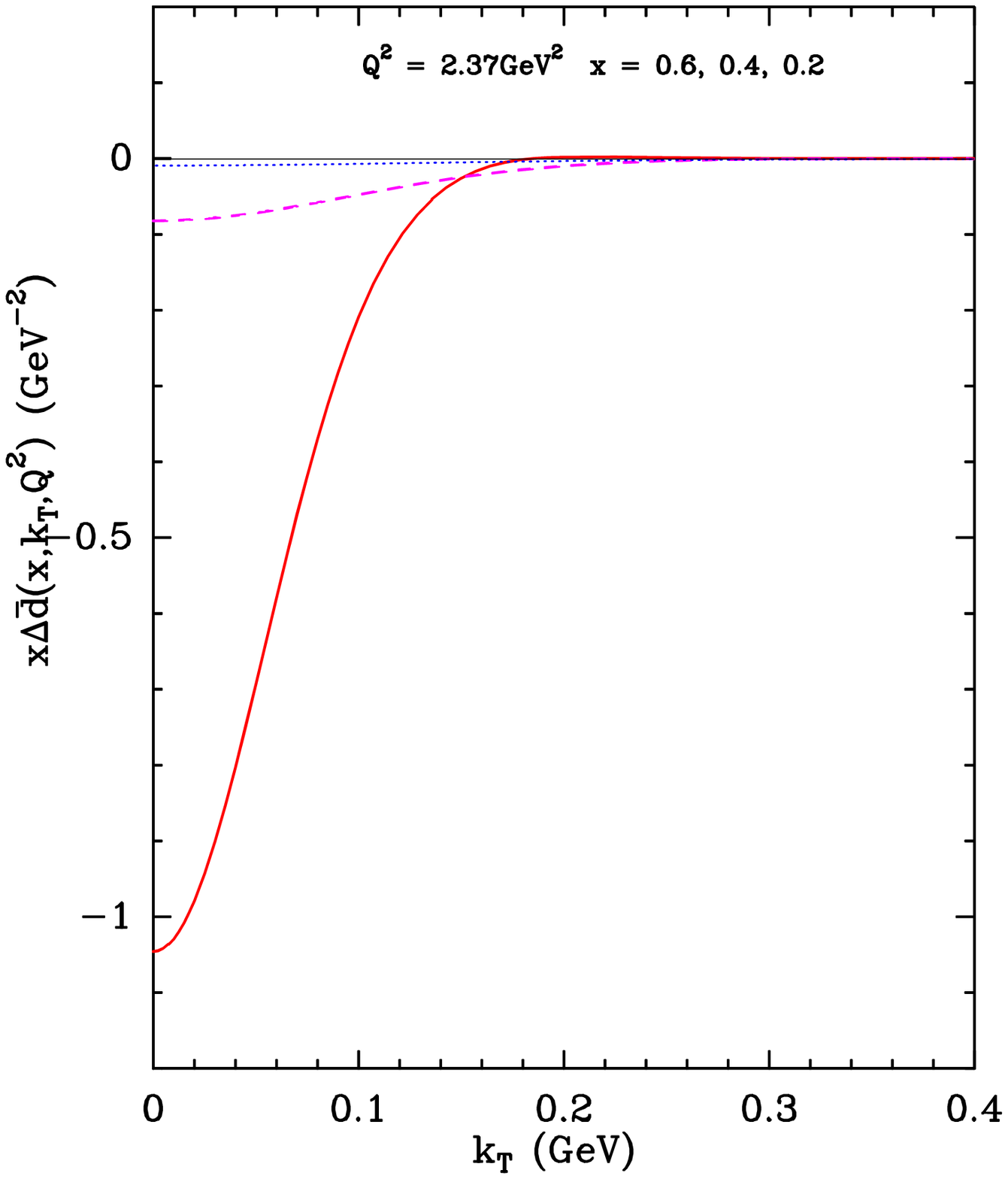,width=8.2cm}
    \end{minipage}
\end{center}
  \vspace*{-10mm}
\caption{
The relativistic covariant distributions $x\Delta \bar u(x,k_T,Q^2)$ ($\it left$)
and $x\Delta \bar d(x,k_T,Q^2)$ ($\it right$), calculated at
$Q^{2}=2.37~\mbox{GeV}^2$, versus $k_T$, for different $x$ values: solid line
$x=0.2$, dashed line $x=0.4$, dotted line $x=0.6$.}
\label{zavdelbarud}
\vspace*{-1.5ex}
\end{figure}

\begin{figure}[htbp]
\begin{center}
\includegraphics[width=6.7cm]{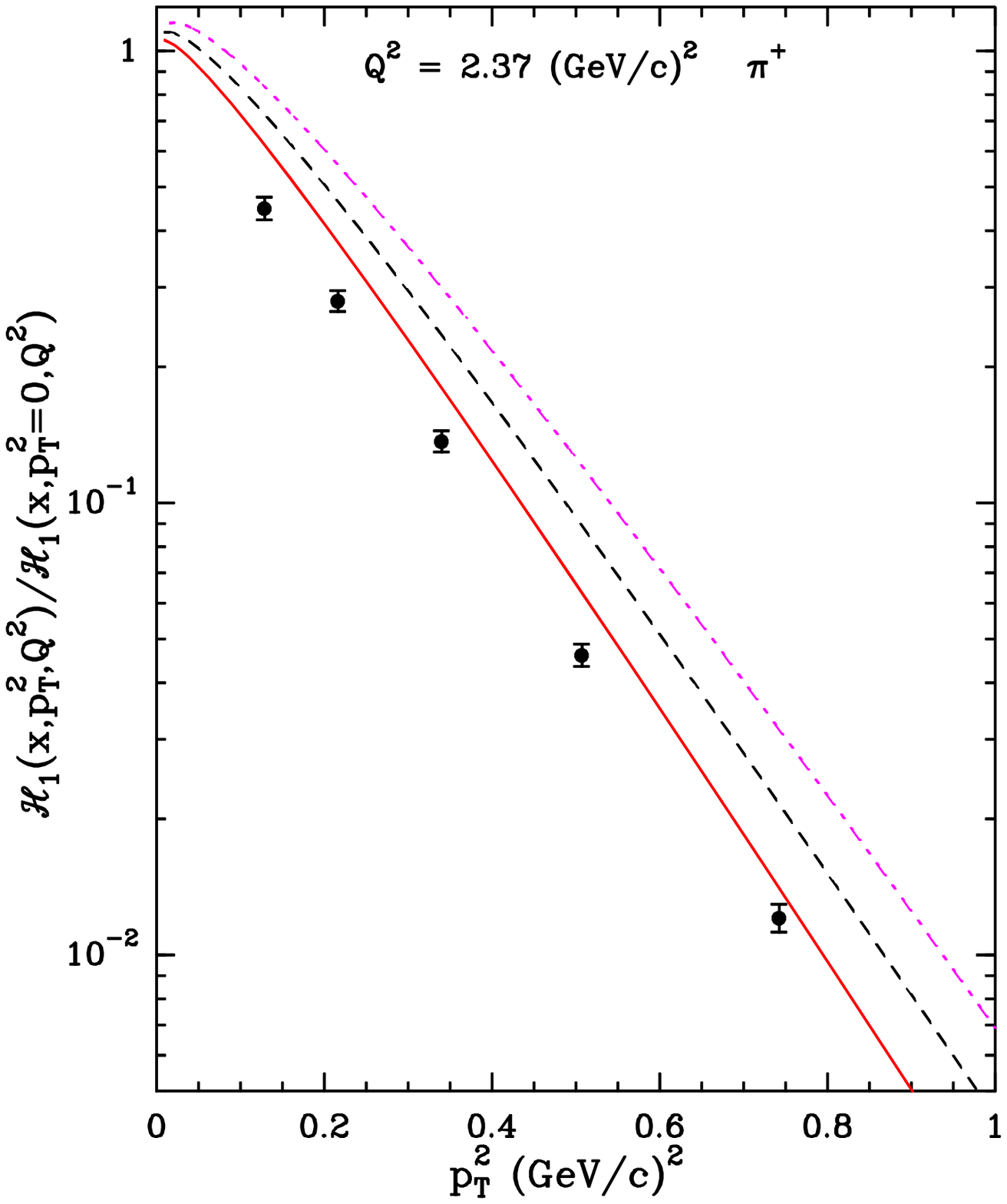}
\includegraphics[width=6.7cm]{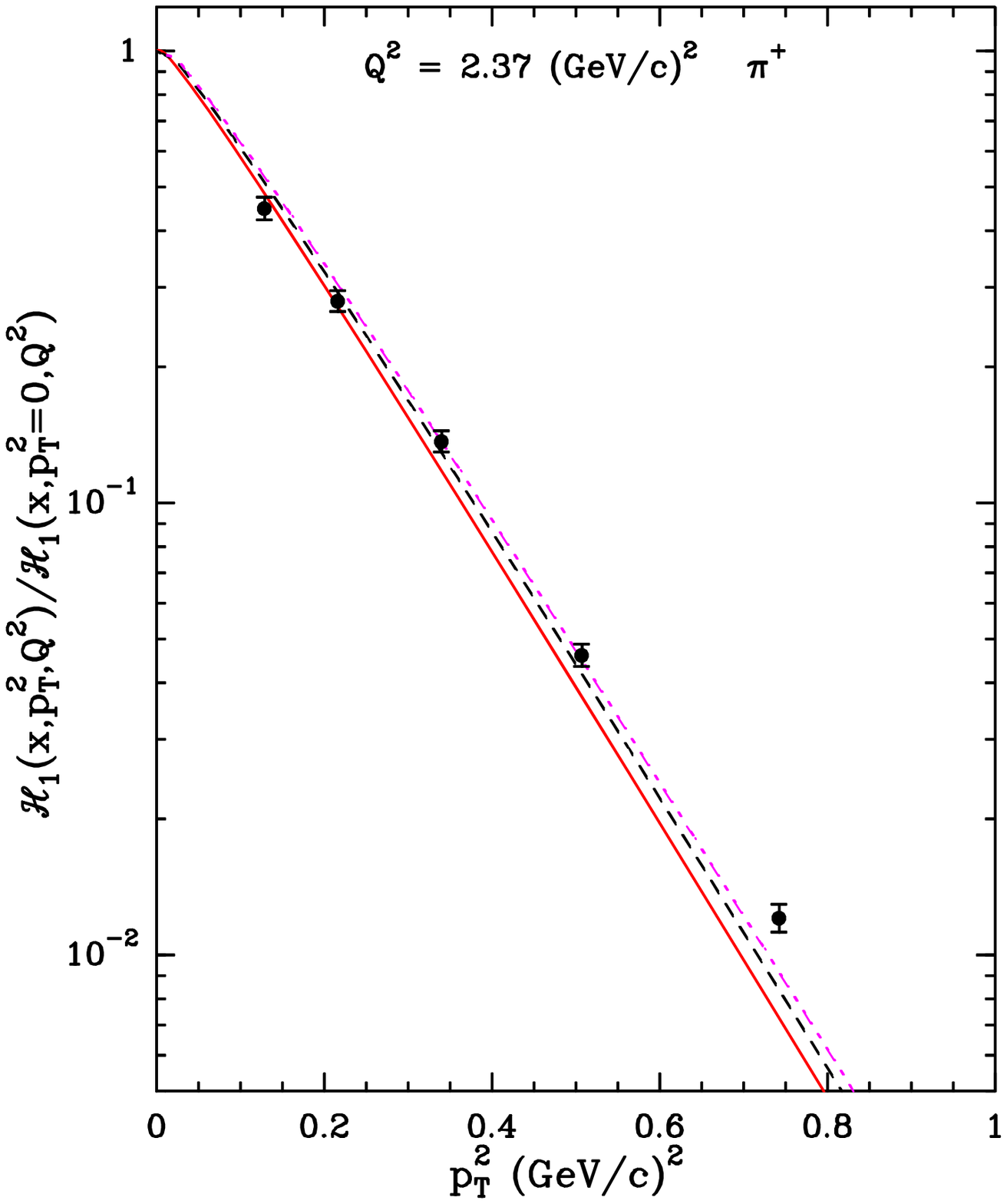}
\caption[*]{\baselineskip 1pt
The $p_T^2$ dependence of the term $\mathcal{H}_1$ at $Q^2= 2.37\mbox{GeV}^2$
and $z=0.30$ for $\pi^+$ production on a proton target. Comparison
of the results of the statistical approach ($\it left$) and the relativistic
covariant distributions ($\it right$). In both cases the solid lines are
for $x=0.20$, the dashed lines for $x=0.40$ and the dotted lines for $x=0.60$. The data are from Ref.~\cite{clas1}
for $x=0.24$ and the error bars are statistical only.}
\label{crosszav1}
\end{center}
\end{figure}

\begin{figure}[htp]
\vspace*{-3.5ex}
\begin{center}
\includegraphics[width=9.0cm]{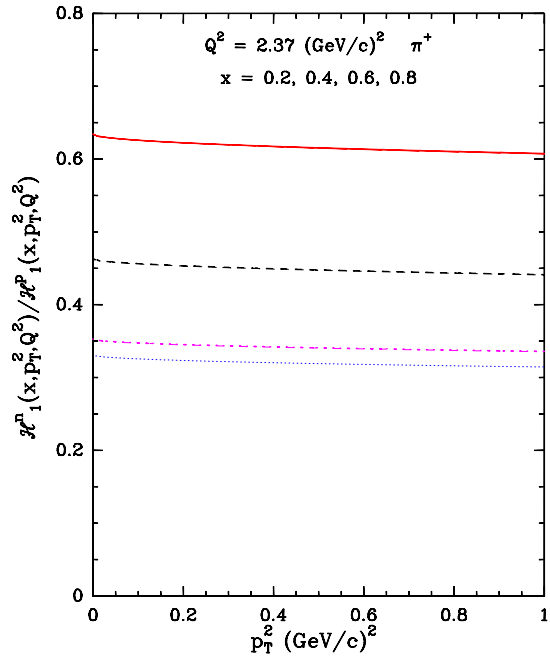}
\caption[*]{\baselineskip 1pt
The $p_T^2$ dependence of the ratio $\mathcal{H}^n_1/\mathcal{H}^p_1$
 at $Q^2= 2.37\mbox{GeV}^2$
and $z=0.30$ for $\pi^+$ production on a neutron and proton target.
Solid line is
for $x=0.20$, dashed line for $x=0.40$, dashed dotted line 
for $x=0.60$ and dotted line for $x=0.80$ .}
\label{rapnp}
\end{center}
\end{figure}

\begin{figure}[htbp]
  \vspace*{-30mm}
\begin{center}
  \vspace*{-15mm}
\includegraphics[width=6.7cm]{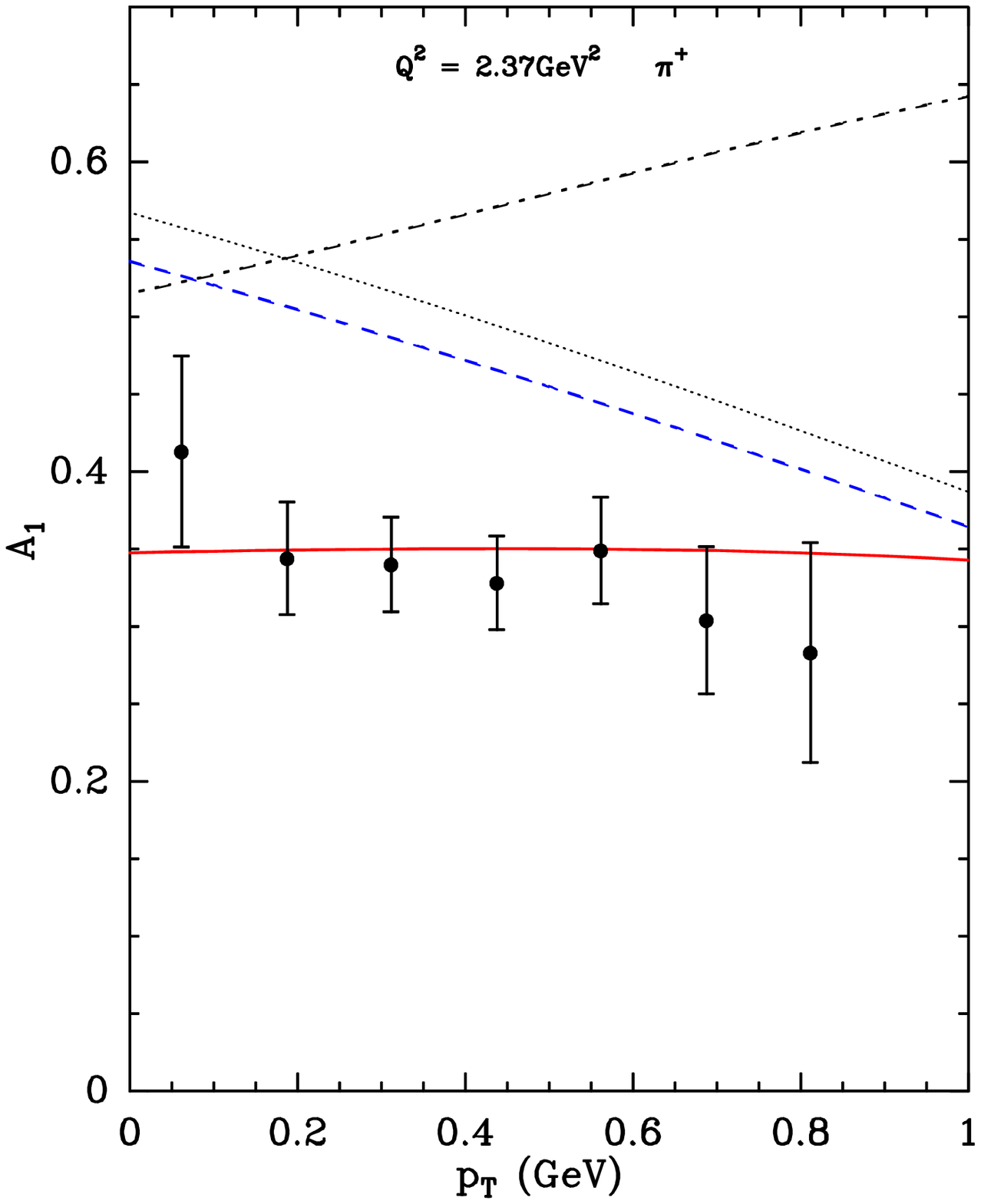}
\includegraphics[width=6.7cm]{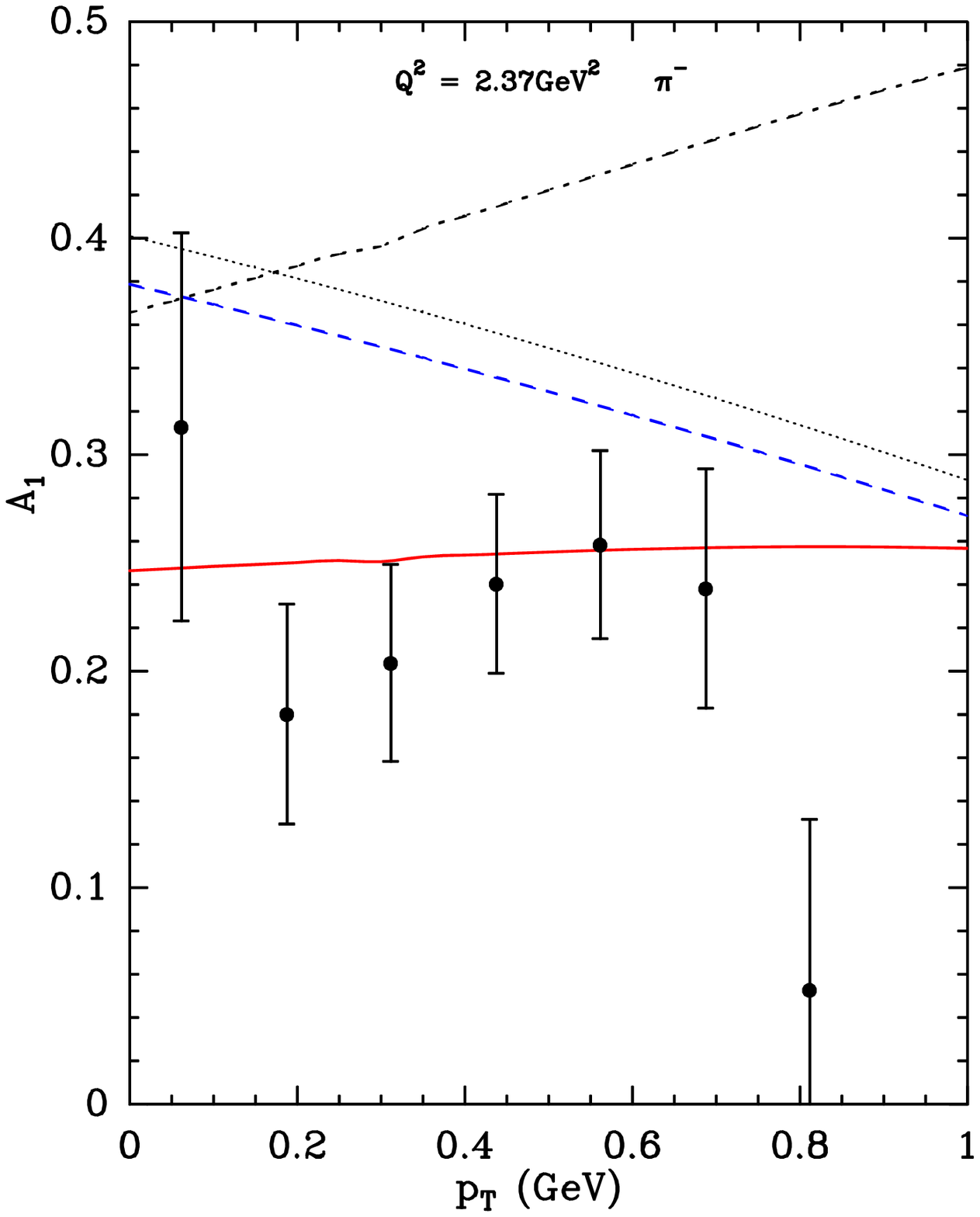}
     \begin{minipage}[t]{0.300\textwidth}
  \vspace*{-25mm}
\centerline{
      \includegraphics[width=1.7\textwidth,height=0.5\textheight]{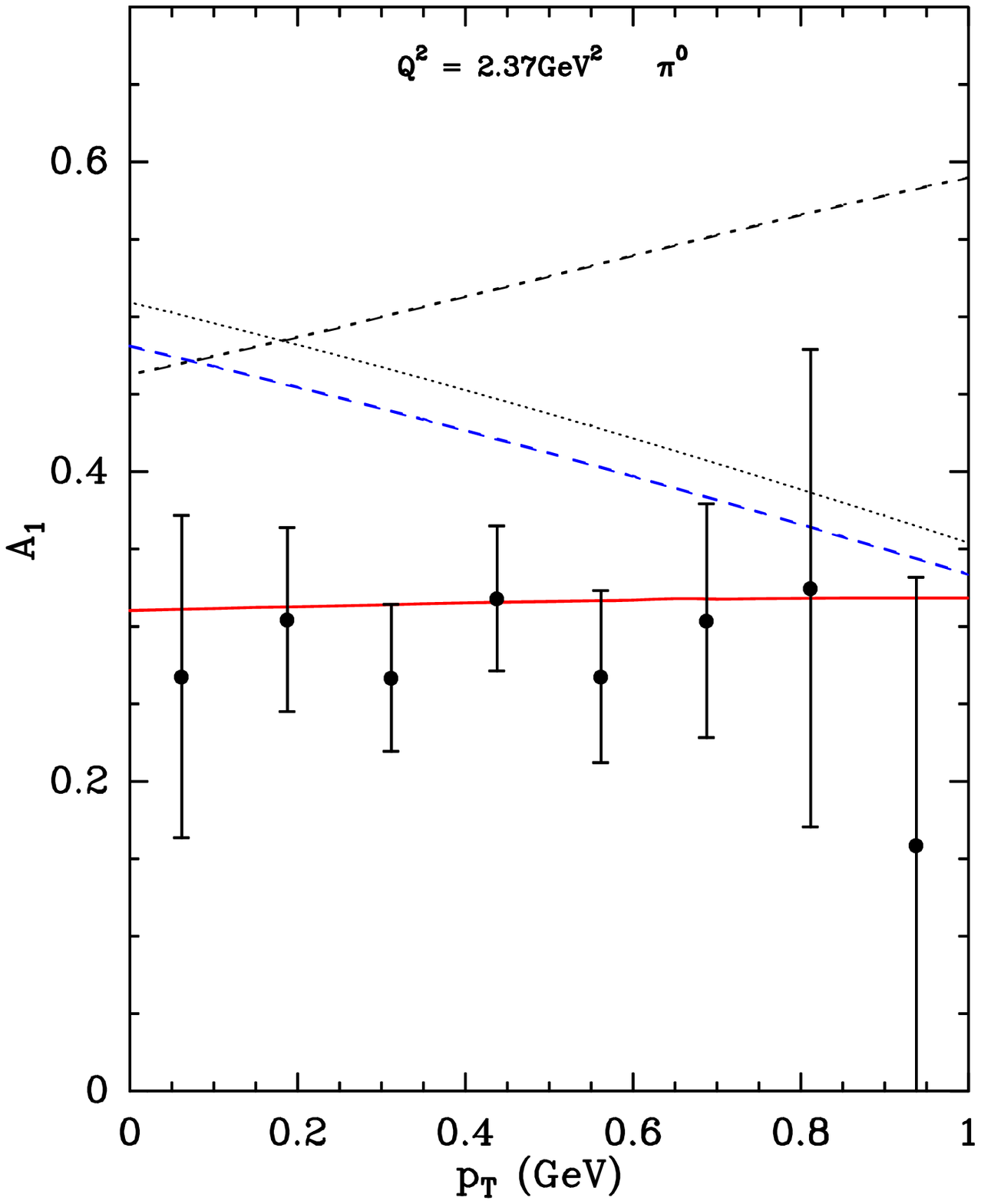}}
    \end{minipage}
\end{center}
\vspace{-5.0mm}
\caption{
The double longitudinal-spin asymmetry $A_1$ for $\pi^+$ ($\it {top-left}$), $\pi^-$ ($\it {top-right}$) and
$\pi^0$ ($\it bottom$) production on a proton target, versus $p_T$, with
the following kinematic cuts corresponding to the JLab data Ref.~\cite{clas2}:
$0.12<x<0.48$, $0.4<y<0.85$ and $0.4<z<0.7$. The data displayed are those of Ref.~\cite{clas2} and the error bars are statistical only. The solid lines are the
results from the statistical distributions and the dashed dotted lines without the inclusion of the Melosh-Wigner rotation. The dashed lines
correspond to the relativistic covariant distributions and the dotted lines without the inclusion of the Melosh-Wigner rotation.}
\label{A1ppi}
\end{figure}

\begin{figure}[htbp]
\begin{center}
\includegraphics[width=6.5cm]{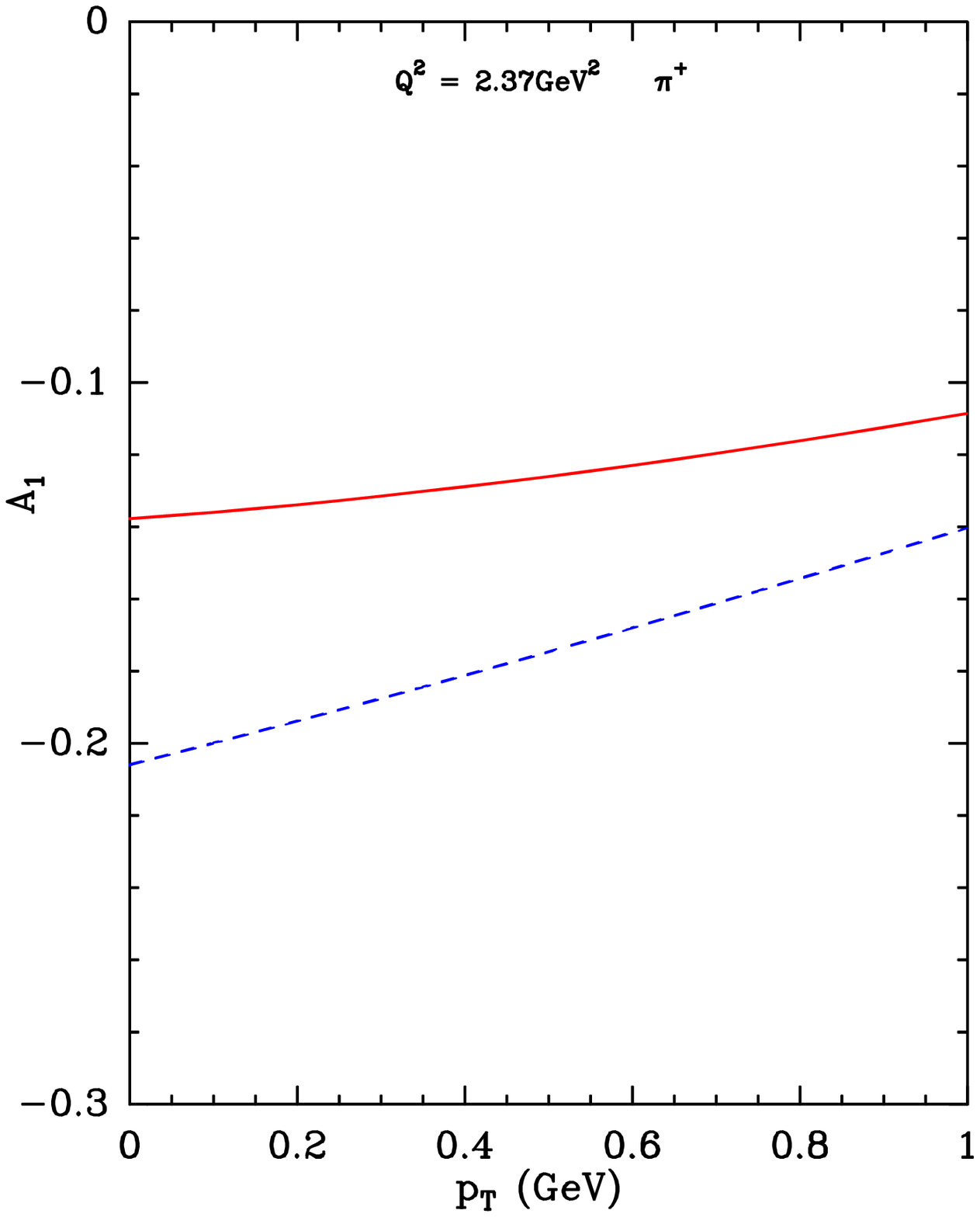}
\includegraphics[width=6.5cm]{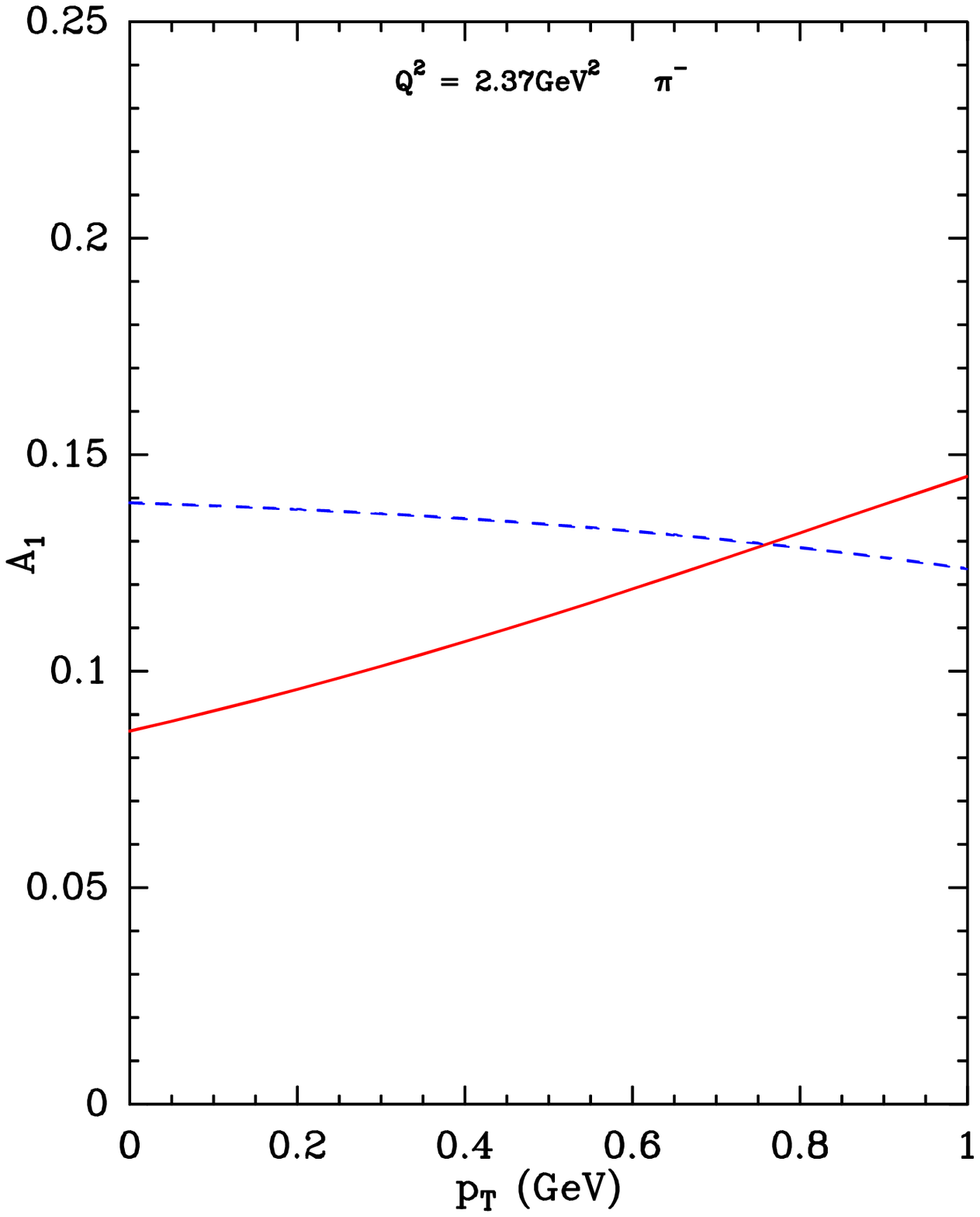}
\caption[*]{\baselineskip 1pt
The double longitudinal-spin asymmetry $A_1$ for $\pi^+$ ($\it left$) and
$\pi^-$ ($\it right$) production on a neutron target, versus $p_T$, assuming
the same kinematic cuts as for the proton target. The solid lines are the
results from the statistical distributions and the dashed lines correspond to
the relativistic covariant distributions.}
\label{crosszav2}
\end{center}
\end{figure}

In this section we will present all our results on the TMD unpolarized and
polarized PDF, for light quarks and antiquarks, resulting from
the two approaches considered above. We will discuss their specific features
and the difference they lead to, in the calculation of the cross sections and
the spin asymmetries for SIDIS pion production. All our results are given at $Q^2=2.37 \mbox{GeV}^2$, the value corresponding to the CLAS data \cite{clas1,clas2}, so we have performed a backward QCD evolution from our input energy scale $Q_0^2=4 \mbox{GeV}^2$.\\
In Fig.~\ref{bbsud} we show $xu(x,k_T,Q^2)$ and $xd(x,k_T,Q^2)$ 
as a function of $k_T$
for different $x$ values, using the TMD statistical PDF constructed in section
\ref{tmdrev}. We have checked that in this $x$-region, the non-diffractive part 
of the quark distributions largely dominate. Similarly $x\Delta u(x,k_T,Q^2)$ and
$x\Delta d(x,k_T,Q^2)$ are shown in Fig.~\ref{bbsdelud} and we recall 
that they are
independent of the diffractive contribution. It is clear that all these $k_T$
distributions are close to a Gaussian behavior, but with  a $x$-dependent
width. The corresponding antiquark PDF are shown in Figs.~\ref{bbsbarud}
and \ref{bbsbardelud} and in this
case, we notice a much rapid fall off in $k_T$ compare to
the unpolarized PDF.\\
Now if one uses the procedure resulting from the relativistic covariant
approach described in section \ref{tmdcov}, one obtains different TMD unpolarized and
polarized PDF as shown in Figs.~\ref{zavud} and \ref{zavdelud}
for $u$ and $d$ quarks. By comparing
them with Figs.~\ref{bbsud} and \ref{bbsdelud}, 
we see that their $k_T$ fall off is much faster than
in the previous case. The corresponding antiquark distributions 
are shown in Figs.~\ref{zavbarud} and \ref{zavdelbarud}.\\
 Next, we turn to the calculation of the unpolarized cross section and the double
longitudinal-spin asymmetry for pion production in polarized SIDIS. The cross
section is directly related to ${\cal H}_1$ (see Eq.~(\ref{2terms})) and we
show in Fig.~\ref{crosszav1}, 
the results, on a proton target, from the two approaches,
versus $p_T^2$, for different $x$-values. The parameter $\mu_D^2$, which enters in the fragmentation function, Eq.~(\ref{dfffg2}), is a free parameter. In
 order to get the best description of the data of Ref.\cite{clas1}, it has been adjusted to the value $\mu^2_D=0.155\mbox{GeV}^2$, 
a value slightly different from the one used in Ref.~\cite{aekp}.
 The agreement is better in the case of the relativistic covariant
approach, which has a faster $p_T^2$ fall off. However it has a very little $x$-dependence, which is 
not known experimentally at the moment, unfortunately. We
predict essentially the same result for $\pi^-$ production and also for
$\pi^{\pm}$ production on a neutron target.\\
However if we consider the ratio of the cross sections for the production of 
a $\pi^+$ from
a neutron target over its production from a proton target, our prediction is 
shown in Fig. \ref{rapnp}. The calculation was done in the two approaches considered and the results are
almost identical. At fixed $p_T$, this 
ratio decreases with $x$, following almost the trend of the ratio $d(x)/u(x)$ 
(see Fig. 4 of Ref.~\cite{bbs4}) and at fixed $x$, it is essentially flat over 
$p_T$, because the TMD of the pion fragmentation function is flavor 
independent. This prediction is worthwhile to check with future experiments.\\
Finally, let us consider the double longitudinal-spin asymmetry $A_1$, defined
above. The results of the calculation from both approaches, for $\pi^{\pm,0}$ on
a proton target are shown in Fig.~\ref{A1ppi}, with the kinematic cuts 
corresponding to
the JLab recent data \cite{clas2}. Whereas the relativistic covariant approach
leads to an asymmetry decreasing with $p_T$, the statistical approach leads to
a flat dependence in $p_T$, in fairly good agreement with the data, and it 
gives the correct
normalization. We note that this behavior, which was obtained with
$\kappa=1.35$ in (\ref{mel3}), is partly due to the effect of the
Melosh-Wigner rotation. In both approaches this effect reduces $A_1$, but it plays an essential role 
in the statistical approach because it compensates the
effect of the $k_T$ rising behavior of $\Delta q$. For completeness, in
view of future experiments, we have also calculated the asymmetry on a neutron
target, which are displayed in Fig.~\ref{crosszav2}. For the production of 
$\pi^+$, $A_1$ is
sensitive to $\Delta d$ and this is the reason for a negative result. In this
case, the predictions from the two approaches lead again to rather different
results, which should be also compared to the predictions from
Ref.~\cite{aekp}.\\
Before closing this discussion we must come back to the effect of the
Melosh-Wigner
rotation. It is clear that the integral over $k_T$ of $\Delta q'$ (see
Eq.~(\ref{mel3})
is smaller than $\Delta q(x)$. Therefore to solve this small mismatch, one
should adjust
the potentials, in such a way that $X_{0q}^+ -  X_{0q}^-$ increases slightly,
whereas $X_{0q}^+ + X_{0q}^-$ remains unchanged, since the
Melosh-Wigner rotation does not affect the unpolarized distribution $q$. This
new improvement will be considered more seriously in future work, when
 we will have access to more precise data on the $k_T$ dependence of the
quark distributions, allowing also a good flavor separation between $u$ and
$d$ quarks.

\section{Concluding remarks}
An important result of this work is the construction of a new set of TMD
statistical distributions. This allows us to take into account, in a
satisfactory way, the multiplicative factors $X^{h}_{0q}$ and
$(X^{-h}_{0q})^{-1}$ in the numerators of the non-diffractive parts of $q$'s
and $\bar{q}$'s distributions. We have introduced some thermodynamical
potentials $Y^h_{0q}$, associated to the quark transverse momentum $k_T$, and
related to $X^{h}_{0q}$ by the simple relation
$\mbox{ln}(1+\exp[Y^h_{0q}])=kX^h_{0q}$. This approach involves a parameter
$\mu^2$, which plays the role of the temperature for the transverse degrees of
freedom and whose value was determined by the transverse energy sum rule. The
substitution $x \to x^2$, we had to make in the diffractive part of the quark
(antiquark) distributions and in the gluon distribution \footnote{ Let us
recall that, unlike the non-diffractive part, they didn't require any
multiplicative factor.}, to avoid a singularity in the energy sum rule, remains
an open problem at the moment. We can give the intuitive argument that gluons,
as photons in a laser, are created mainly in the forward direction. The
diffractive part, which comes from their conversion into $q\bar q$ pairs, as
well as the gluons themselves, may thermalize only for the $x$ degree of
freedom.\\
We have calculated the $p_T$ dependence of SIDIS cross sections and double
longitudinal-spin asymmetries, taking into account the effects of the
Melosh-Wigner rotation, for $\pi^{\pm,0}$ production by using this
 set of TMD parton distributions and another set coming from the relativistic
covariant approach. These sets lead to different results, which were compared
to recent experimental data. Both sets do not satisfy the usual factorization
assumption of the dependence in $x$ and $k_T$. We have made some predictions
for future experiments with neutron targets, which will allow futher tests of
our results. Major progress in our understanding of the TMD PDF will be certainly
achieved also by measurements at a future electron ion collider \cite{duke}.\\

\appendix
\section{Appendix}
\setcounter{equation}{0}
\numberwithin{equation}{section}

Let us consider the contribution of the diffractive term to the energy sum
rule. So by using Eq.~(\ref{diff}), this contribution reads
\begin{equation}
\int\frac{k_T^2}{x} q^D(x , k_T^2) dx dk_T^2  =
\int_0^1 \frac{\tilde A x^{\tilde b - 2}}{\mu^2 x^2\mbox{ln}2 (\exp{\frac{x
}{\bar{x}}} + 1)} dx
\int_0^\infty \frac{k_T^2 d k_T^2}{\exp{(\frac{k_T^2}{ x^2 \mu^2})} + 1}~,
\label{A1}
\end{equation}
and after the change of variable $\xi=k_T^2/x^2 \mu^2$, we get
\begin{equation}
\mu^2 \int_0^1\frac{\tilde A x^{\tilde b} dx}{\exp{\frac{x}{\bar{x}} + 1}}
\int_0^\infty\frac{
\xi d \xi}{\exp{\xi} + 1} = \frac{\mu^2 \pi^2P_D}{ 12 \ln{2}}
\label{A2}
\end{equation}
One finds that $P_D=0.0115$, so the contribution of the diffractive part to the
sum rule is negligible.\\
Finally let us consider the case of the gluon. In order to recover its
$x$-dependence (see Eq.~(\ref{eq5})), we can write
\begin{equation}
x G(x,k_T^2) = - \frac{A_G x^{b_G-2}}{(\exp{\frac{x}{\bar{x}}} - 1)\mu^2 \ln(1
- \exp Y_G)}
\frac{1}{\exp{(\frac{k_T^2}{x^2 \mu^2}+ Y_G)} - 1}~,
\label{A3}
\end{equation}
because we had to introduce a small $Y_G$, otherwise $x G(x,k_T^2)=0$. The
contribution to the energy sum rule Eq.~(\ref{esr}) is
\begin{equation}
\int\frac{k_T^2}{x} G (x , k_T^2) dx dk_T^2 = -
\int_0^1 \frac{A_G x^{b_G-4}}{\mu^2 \ln(1 - \exp Y_G)(\exp{\frac{x}{\bar{x}}} -
1)} dx
\int_0^\infty \frac{k_T^2 d k_T^2}{\exp{(\frac{k_T^2}{ x^2 \mu^2}+ Y_G)} - 1}~,
\label{A4}
\end{equation}
and after the change of variable $\xi=k_T^2/x^2\mu^2$, we get
\begin{equation}
- \mu^2 \int_0^1\frac{A_G x^{b_G} dx}{(\exp{\frac{x}{\bar{x}}-1})\ln(1 - \exp
Y_G)} \int_0^\infty\frac{
\xi d \xi}{\exp({\xi + Y_G}) - 1} = \mu^2 P_G \cdot I_G~,
\label{A5}
\end{equation}
with
\begin{equation}
P_G=\int_0^1\frac{A_G x^{b_G} dx}{\exp{\frac{x}{\bar x}}-1} = 0.421
\label{A6}
\end{equation}
and
\begin{equation}
I_G= - \frac{1}{\ln(1 - \exp Y_G)} \int_0^\infty\frac{
\xi d \xi}{\exp({\xi + Y_G}) - 1} = - \frac{\mbox{Li}_2(\exp Y_G)}{\ln(1 -
\exp Y_G)}~.
\label{A7}
\end{equation}
With $Y_G = 10^{-6}$, we find $I_G = 0.119$, so the contribution of the gluon
to the energy sum rule is negligible.

\clearpage
\newpage
\clearpage
\newpage
\clearpage
\newpage
\clearpage
\newpage

\clearpage
\newpage

\clearpage
\newpage

\clearpage
\newpage

\clearpage
\newpage

\clearpage
\newpage

\clearpage
\newpage

\clearpage
\newpage

\clearpage
\newpage
\end{document}